\documentclass{CVM}

\usepackage{amsmath,amsfonts}
\usepackage{algorithmic}

\usepackage{array}
\usepackage{pgfplots}

\usepackage{textcomp}
\usepackage{stfloats}
\usepackage{url}
\usepackage{verbatim}
\usepackage{graphicx}
\usepackage{makecell}
\usepackage{booktabs}
\usepackage{multirow}
\usepackage{float}
\usepackage{xcolor}

\CVMsetup{
	type      = {Research Article},
	doi       = {s41095-0xx-xxxx-x},
	title     = {Super-resolution Reconstruction of Single Image for Latent features},
	author    = {Xin Wang$^{1}$$^,$ $^{3}$$^,$ $^{4}$,  Jing-Ke Yan$^{2}$$^,$$^{3}$ \cor{}, Jing-Ye Cai$^{1}$\cor{}, Jian-Hua Deng$^{1}$,  Qin Qin$^{3}$, Yao Cheng$^{2}$                 },
	runauthor = { Jing-Ke Yan, Xin Wang, Jing-Ye Cai},
	abstract  = {
		Single-image super-resolution (SISR) typically focuses on restoring various degraded low-resolution (LR) images to a single high-resolution (HR) image. However, during SISR tasks, it is often challenging for models to simultaneously maintain high quality and rapid sampling while preserving diversity in details and texture features. This challenge can lead to issues such as model collapse, lack of rich details and texture features in the reconstructed HR images, and excessive time consumption for model sampling. To address these problems, this paper proposes a Latent Feature-oriented Diffusion Probability Model (LDDPM). First, we designed a conditional encoder capable of effectively encoding LR images, reducing the solution space for model image reconstruction and thereby improving the quality of the reconstructed images. We then employed a normalized flow and multimodal adversarial training, learning from complex multimodal distributions, to model the denoising distribution. Doing so boosts the generative modeling capabilities within a minimal number of sampling steps. Experimental comparisons of our proposed model with existing SISR methods on mainstream datasets demonstrate that our model reconstructs more realistic HR images and achieves better performance on multiple evaluation metrics, providing a fresh perspective for tackling SISR tasks.
	},
	keywords  = {Image Superresolution Reconstruction;  Denoising Diffusion Probabilistic Model; Normalized Flow; Adversarial Neural Network; Variational Auto-Encoder},
	copyright = {The Author(s) 2023. },
}

\begin{document}

\maketitle
\begin{figure}[hb]  \vskip -4mm
	\renewcommand\arraystretch{1}
	\begin{tabular}{p{80.5mm}} \toprule\\ \end{tabular}
	\vskip -4.5mm \noindent \setlength{\tabcolsep}{1pt}
	\begin{tabular}{p{3.5mm}p{80mm}}
		
		$1\quad $ & School of Information and Software Engineering, University of Electronic Science and Technology of China, Chengdu, 610000,  China.  E-mail: J. Cai, jycai@uestc.edu.cn \cor{} ; J. Deng, Jianhua.deng@uestc.edu.cn.com;  X. Wang, 303479506@qq.com.\\
		$2\quad $ & State Key Laboratory of Rail Transit Vehicle System, Southwest Jiaotong University, Chengdu, 610000, China. E-mail: J. Yan, 592499985@qq.com \cor{} , Y. Cheng, chengyao2020@swjtu.edu.cn).\\
		$3\quad $ & School of Marine Engineering, Guilin University of Electronic Technology, Guilin, 541000, China. E-mail: Q. Qin, qinqin@mails.guet.edu.cn.\\
		$4\quad $ & School of Computer and Information Security, Guilin University of Electronic Technology, Guilin, 541000,  China. \\

		&\hspace{-5mm} Manuscript received: 2023-00-00; accepted: 2023-00-00 \vspace{-2mm}
	\end{tabular} \vspace {-3mm}
\end{figure}

\section{Introduction}
Single-image super-resolution (SISR) holds great significance in research fields such as computer vision \cite{DBLP:journals/tip/ChengFLZSW22}\cite{jiang2023sphere} and image processing \cite{DBLP:journals/tii/WangXLXX22}\cite{chen2023towards}. SISR tasks involve the reconstruction of high-resolution (HR) images from corresponding low-resolution (LR) images degraded by various factors. Owing to the loss of a substantial amount of detail and texture features in LR images during multiple degradation processes, the reconstructed HR images often lack rich image details and clear textures. Therefore, the SISR task is a typical problem for addressing uncertainty. In the context of SISR tasks, researchers have successively proposed various traditional methods, such as iterative backprojection \cite{DBLP:journals/vc/MaT21}, convex set projection \cite{DBLP:journals/spic/KarimiT21}, and sparse representation \cite{DBLP:conf/cvpr/ZhouHGZ21}. However, traditional methods generally involve explicit estimation of the blur kernel before reconstructing the HR image, leading to errors in the estimated blur kernel and, consequently, suboptimal HR image reconstruction results.

The SISR task can also be viewed as a common generative task. This type of task aims to fit the data distribution of reconstructed images effectively through a generator, bringing the data distribution of the reconstructed images as close as possible to that of the real images. Deep learning-based methods in this task can be divided into five categories: regression-based, generative adversarial network (GAN)-based \cite{DBLP:journals/tgrs/ShiHHCHD22}, flow-based \cite{DBLP:conf/iccv/LiangL0DGT21}, variational auto-encoder (VAE)-based \cite{DBLP:conf/cvpr/LiuSW21}, and denoising diffusion probabilistic model (DDPM)-based \cite{DBLP:journals/ijon/LiYCCFXLC22} methods. However, generative models face three major challenges in SISR tasks: high-quality sampling, fast sampling, and maintaining detail diversity after sampling.
Regression-based methods can fit any function but not any data distribution. Therefore, solving problems such as unrealistic perception and artifacts in the reconstruction results using regression-based methods alone is challenging. GAN-based methods are also commonly used for SISR tasks. They leverage perceptual and adversarial losses for image reconstruction. Although these methods can provide fast sampling, they are prone to issues such as model collapse and unstable training. Flow-based methods use a series of reversible transformations and sample the log-likelihood function to infer latent variables accurately, thereby improving the diversity of the generated images. The drawback is that the generated images are overly smooth and incur significant computational overhead. VAE-based methods can not only generate more diverse data using additional conditions but also offer relatively fast sampling. However, the quality of sampling in VAE-based methods tends to be low, leading to a loss of detail and texture in the HR images.

Recently, DDPM has achieved commendable results in generative tasks, such as image synthesis \cite{DBLP:journals/corr/abs-2111-14822} and speech synthesis \cite{DBLP:conf/icml/KimKS21}. DDPM utilizes a Markov chain to transform latent variables in a Gaussian distribution into complex data distributions, thereby addressing the "one-to-many" problem in SISR tasks and improving the quality of reconstructed data. However, applying DDPM to SISR tasks presents unique challenges because these tasks differ from other generative tasks as follows:

\begin{enumerate}
	\item{In SISR tasks, the reverse diffusion of DDPM requires a complex probability distribution to model the denoising distribution. Consequently, the forward diffusion of DDPM requires thousands of evaluation steps to sample a single feature. If DDPM uses a limited number of sampling steps, the quality of the image generated after sampling is low.}
	
	\item{DDPM is based on unconditional or simple conditions for model input. On the contrary, SISR tasks often require the full utilization of LR images as conditions for model input to constrain the solution space for HR images.}
\end{enumerate}

Therefore, this paper proposes a novel approach called the Latent Feature-oriented Diffusion Probability Model (LDDPM) to overcome the challenges faced by DDPM in SISR tasks. Specifically:
\begin{enumerate}
	\item{To extract feature information from LR images to constrain the solution space for HR images, we designed an adaptive multi-head attention mechanism and a variational autoencoder to transform LR images into latent conditions for model input.}
	
	\item{To ensure that the DDPM can perform high-quality sampling with a limited number of steps, we designed a multimodal distribution for modeling HR images. This distribution is implemented based on GANs and normalized flows, enabling LDDPM to focus on reconstructing high-frequency details of HR images with fewer iterative steps.}
	
\end{enumerate}

Our model has the following advantages:

\textbf{Stable Style and Content Consistency:} Although the probability distribution of HR images is difficult to predict, we designed a novel conditional encoder to limit the impact of prediction randomness introduced by maximizing the variational lower bound in DDPM. This stabilizes the model training and ensures the generation of images that are consistent with the original HR image in terms of style and content.

\textbf{Fast and High-Quality Sampling:} By leveraging Markov chains and modeling HR images using a complex multimodal distribution, our model can sample rapidly while mitigating the negative impacts of model collapse on the reconstruction of HR images. Consequently, it can generate diverse and high-quality HR images

Our LDDPM was empirically validated on numerous datasets. The experimental results demonstrate that our model outperforms most methods in SISR tasks across multiple datasets. 
\section{Related Work}
\begin{figure}[H]
	\centering
	\includegraphics[width=\linewidth]{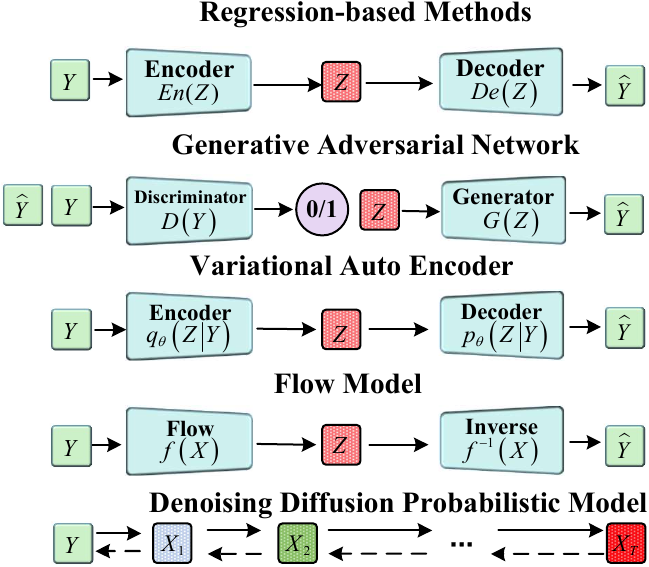}
	\caption{Mainstream generative models}
	\label{fig_1}
\end{figure}

In this section, we discuss methods that employ generative models, including regression-, GAN-, flow-, VAE-, and DDPM-based methods, as illustrated in Figure \ref{fig_1}.

\subsection{Single-image Super-resolution Reconstruction Based on Traditional Generation Model}
\subsubsection{Regression-based methods}
Owing to the rapid development of deep learning, many deep learning-based methods have been proposed for SISR tasks. Most are regression-based methods that leverage end-to-end learning to map the relationship between LR and HR images. For instance, Zhang et al. \cite{DBLP:journals/tcsv/ZhangLWPMYY22} noted that most regression-based methods do not adequately explore contextual information in LR images during feature extraction. Therefore, they proposed a two-stage single-image reconstruction method (TSAN)-based attention mechanism, which accurately reconstructs HR images in a coarse-to-fine manner. However, TSAN rarely exploits the interlayer feature correlation, thereby reducing the ability of convolutional neural networks to learn the feature mapping relationship between LR and HR images. Dai et al. \cite{DBLP:conf/cvpr/DaiCZXZ19} addressed this issue by capturing long-distance spatial contextual information between features using a second-order feature statistics module and a non-local enhanced residual module, enabling the model to learn abstract feature mapping representations. Although the second-order feature statistics module effectively extracts information-rich features at each layer, it processes the features of each convolutional layer independently. This approach overlooks the correlation between features at different layers. Consequently, Niu et al. \cite{DBLP:conf/eccv/NiuWRZYWZCS20} introduced a regression-based holistic attention network (HAN) that not only considers interlayer correlations and adaptively emphasizes interlayer features but also learns the confidence of each channel, enabling the model to map complex feature relationships. In SISR tasks, similar patches in images can provide information to each other, helping the model learn the feature mapping relationship between LR and HR images. Accordingly, Zhou et al. \cite{DBLP:conf/nips/ZhouZZL20} divided LR images into multiple blocks and used each block to search for the $K$ nearest neighboring features to dynamically build a cross-scale matrix. This approach allows the feature relationships in HR images to be passed to the query patch in LR images, aiding the recovery of more complex details and texture features. However, HR images generated using regression-based methods alone may appear perceptually unrealistic and contain artifacts. A breakthrough solution to this problem is the use of GAN-based methods.

\subsubsection{Generative adversarial network-based methods} 
GANs utilize generators and discriminators to calculate content and discrimination losses, striving to align the data distribution of the generated images as closely as possible with those of real images. For instance, Wang et al. \cite{DBLP:conf/eccv/WangYWGLDQL18} discovered that if the goal of perceptual loss is to minimize the error in pixel space rather than in feature space, the generated HR images tend to output overly smoothed results, thereby lacking sufficient high-frequency details. In response, they proposed an enhanced super-resolution GAN (SRGAN), which uses activated features to improve perceptual loss, thereby providing a stronger supervisory signal for luminance consistency and texture restoration. However, SRGAN has limited capabilities in reconstructing spectral space invariance, which may lead to spectral distortions in the generated HR images, particularly when the image magnification factor is high. Therefore, Shi et al. \cite{DBLP:journals/tgrs/ShiHHCHD22} mapped the generated spectral features from the image space to the latent space, generating a coupled component to regularize the generated samples. Chan et al. \cite{DBLP:conf/cvpr/ChanWXGL21} designed the Generative LatEnt bANk (GLEAN) model, which significantly reduces the need for costly image-specific optimizations during runtime by directly utilizing a pre-trained GAN and generating magnified images through forward propagation.

Although research advancements in GANs have contributed significantly to the study of SISR tasks, these networks face challenges owing to their data-driven nature. Specifically, they struggle to reconstruct genuine high-frequency information, particularly when dealing with unknown images during testing. Recognizing this limitation, Liu et al. \cite{9815113} integrated the strengths of regression-based methods into GANs owing to their superior adaptability. They applied detailed features captured by convolutional neural networks as prior knowledge to aid GANs in generating more realistic details.

The aforementioned methods produce HR images with fewer artifacts and more realistic perception. However, these methods tend to suffer from model collapse, resulting in ineffective handling of "one-to-many" uncertainties and difficulty in determining the distribution of real samples in the latent space.

\subsubsection{Variational autoencoder-based methods}
VAE-based methods have also been applied to SISR tasks. They initially map the input image to a hidden space to perform probability density estimation. The VAE then assumes a standard Gaussian distribution as the prior and trains a probabilistic decoder to establish a mapping from the latent space to the actual data distribution. For instance, Gatopoulos et al. \cite{DBLP:journals/corr/abs-2006-05218} leveraged the features of neurons in human vision to enhance existing signals by continuously incorporating new information after adapting to light exposure. In their VAE-based methods, they treated downsampled image representations as random variables and continuously fed these variables into the model during training. However, the method proposed by Gatopoulos et al. tends to generate blurry images. To address this issue, Liu et al. \cite{liu2020photo} proposed using a conditional sampling mechanism to narrow the latent subspace for reconstruction. Although Liu's method can reconstruct HR images with simple backgrounds, they used mean square error for model optimization, which could potentially lead to blurring at the edges of images with complex backgrounds. To overcome this problem, Liu et al. \cite{DBLP:conf/cvpr/LiuSW21} proposed searching for images of similar styles from reference images to guide the reconstruction of HR images. They employed a conditional VAE (CVAE) to compress various reference images into a compact latent space, thereby learning an explicit distribution. They then sampled style features from this distribution as conditions or priors to address the edge blurring issue in the reconstructed images. Despite these efforts, the quality of samples generated by VAE-based methods is not optimal, often resulting in the loss of local details and textures.

\subsubsection{Flow-based methods}
Flow-based methods, such as VAE-based methods, are types of generative models. However, flow-based methods employ a bijective function to learn the posterior distribution from the prior distribution through a series of reversible transformation functions, and they generate HR images based on the posterior distribution. For instance, Liang et al. \cite{DBLP:conf/iccv/LiangL0DGT21} found that the normalizing flow model could predict detail-rich HR images from LR images using a joint modeling approach with downsampling and upsampling. Hence, they modeled the LR image and remaining high-frequency components, enabling the model to learn the missing high-frequency information via bijective mapping between HR and LR images. Xiang et al. \cite{xiang2021learning} used a flow-based model for intra-flow feature extraction, inter-flow dependency extraction, and joint feature learning, achieving good HR image reconstruction results. However, these flow-based models are limited by their use of only a few convolutional layers, resulting in a limited receptive field. To address this issue, Jo et al. \cite{DBLP:conf/cvpr/JoYK21} stacked more convolutional layers through affine coupling, which expanded the receptive field and resulted in a stronger feature representation capability. Both flow- and VAE-based methods can effectively learn the distribution of samples in the latent space and solve the "one-to-many" uncertainty issue. However, the detail features of HR images generated by these methods are overly smooth and require a relatively long generation time.

\subsection{Single-image Super-resolution reconstruction Based on Denoising Diffusion Probabilistic Model}
Recently, DDPMs have been used for SISR tasks. DDPMs are composed of two parameterized Markov chains (forward and reverse) and use variational inference to generate samples that are consistent with the original data distribution within finite time. The forward chain perturbs the data by gradually introducing Gaussian noise according to a predetermined noise schedule, until the data distribution approaches the prior distribution (standard Gaussian distribution). The reverse chain begins with a given prior distribution and iteratively learns to restore the original data distribution using a parameterized Gaussian transition kernel.

As such, the DDPM is a highly flexible and computationally efficient generative model that not only effectively avoids the model collapse encountered by GANs but also produces high-quality images. For instance, Lin et al. \cite{DBLP:journals/ijon/LiYCCFXLC22} designed an SRDiff model based on DDPM, which transforms Gaussian noise into HR images through a Markov chain with residuals. Saharia et al. \cite{DBLP:journals/corr/abs-2104-07636} created an image super-resolution model (SR3) based on repetitive refinement, which initially adds Gaussian white noise to the image and then uses the noise image to train the UNet model for iterative refinement of the noise output. Ryu et al. \cite{DBLP:journals/corr/abs-2208-01864} proposed a pyramid denoising diffusion probabilistic model that employs a location embedding-trained score function in addition to the DDPM to gradually generate HR images from LR ones. Xia et al. \cite{DBLP:journals/corr/abs-2303-09472} introduced the DiffIR model, which uses the mapping capability of DDPM to estimate a compact prior representation to guide the reconstruction of HR images, thereby improving the efficiency and stability of DDPM in image restoration. Wang et al. \cite{DBLP:journals/corr/abs-2212-00490} designed a Denoising Diffusion Null-Space Model (DDNM) to solve linear image restoration (IR) issues. This model requires only an existing pre-trained diffusion model as a generative prior, requiring no additional training or network modifications, and can tackle complex reconstruction issues in SISR.

Although the above DDPM-based methods have achieved some results on different super-resolution datasets, the models discussed above directly  stack the noise and LR images at the current moment for conditional sampling without constraining the solution space of the reconstructed HR image. This may lead to inconsistencies in style and content between the reconstructed and original HR images (e.g., an old man wearing glasses may be reconstructed without glasses by the model). Rombach et al. \cite{RombachBLEO22} proposed the LDM model, which not only encodes features in the latent space of HR images using an autoencoder but also encodes LR images using attention mechanisms. Compared with previous works, LDM not only reduces the computational complexity of the model but also constrains the solution space of the reconstructed HR image. However, LDM directly encodes HR images using an autoencoder and trains DDPM using the encoded latent features, which may result in the loss of detail and texture features in the model's reconstructed HR image. Furthermore, although the LDM uses latent features of HR images for image reconstruction, it still requires many diffusion steps to sample HR images. Therefore, the issues of high-quality sampling, diverse samples, and minimal computational overhead encountered by DDPM in SISR tasks are still worth studying.

\section{Methods}
In this section, we present the LDDPM, which is specifically designed for SISR tasks. First, we provide a brief overview of the fundamental architecture of the proposed model. Second, we review the principles of the DDPM. Subsequently, we delve into the key components of the LDDPM. Finally, we introduce the loss function used in the LDDPM.

\subsection{Latent Feature-oriented Denoising Diffusion Probabilistic Model}
\begin{figure*}[ht]
	\centering
	\includegraphics[width=\textwidth]{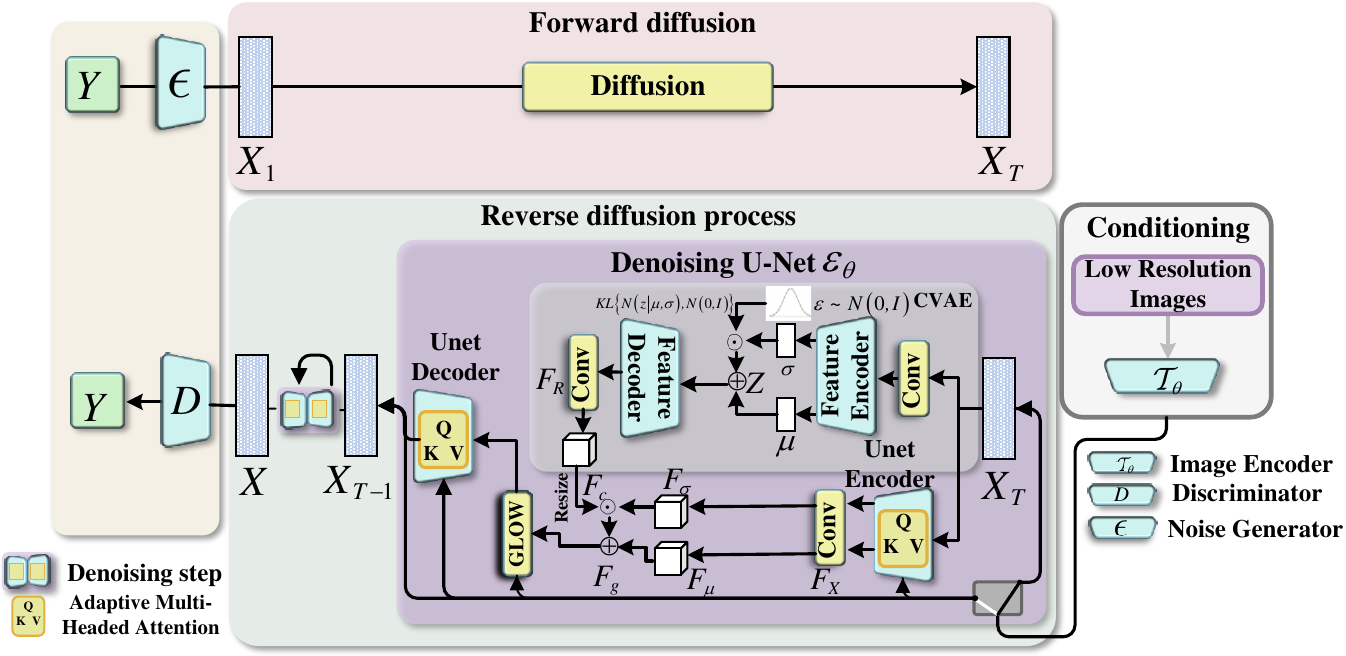}
	\caption{Network architecture of the LDDPM model. Our model can be viewed as a conditional DDPM.}
	\label{fig_2}
\end{figure*}

In SISR tasks, given an LR image $X \in {R^{w \times h \times c}}$, the objective is to recover the corresponding HR image $Y \in {R^{{w_{s \uparrow }} \times {h_{_{s \uparrow }}} \times c}}$. Here, $w$, $h$, and $c$ represent the width, height, and channel number of image $X$, respectively, and $s \uparrow$ represents the upsampling factor. Thus, the problem of SISR tasks can be described by  Eq. (\ref{eq1}):
\begin{equation}
	\begin{array}{l}
		X = k \otimes Y + n
		\label{eq1}
	\end{array}
\end{equation}

where $n$ denotes Gaussian white noise, $k$ denotes the convolution kernel for downsampling, and $\otimes$ denotes the convolutional downsampling. The goal of SISR tasks is to model Eq. (\ref{eq1}) as a maximum a posteriori probability problem, as expressed in Eq. (\ref{eq2}):
\begin{equation}
	\begin{array}{l}
		\widehat {Y} = \mathop {\arg \max }\limits_{Y} \log q\left( {Y} \right) + \log q\left( {{X}\left| {Y} \right.} \right)
		\label{eq2}
	\end{array}
\end{equation}

where $\widehat {Y}$ represents the reconstructed HR image, $\log q({Y})$ refers to the model optimized for HR images, and $\log q({X}|{Y})$ signifies the log-likelihood of LR images given the HR images. However, in traditional SISR tasks, models tend to collapse and struggle to effectively restore image details. DDPM \cite{DBLP:journals/corr/abs-2006-11239} \cite{DBLP:conf/icml/NicholD21} mitigates model collapse and retains additional image details by transforming the standard normal distribution into an empirical data distribution (akin to Langevin dynamics) through a series of refinement steps. Hence, in this work, we employed DDPM's random iterative refinement process to learn the parameters of $\log q({X_1},{X_2}...{X_T}|{Y})$ and thereby approximate $\widehat {Y}$. This process gradually maps multiple source images ${X_1},{X_2}...{X_T}$ to approximate the target image ${Y}$, thus demonstrating a one-to-many mapping. Here, the target image ${Y}$ is made as consistent as possible with multiple source images ${X_1},{X_2}...{X_T}$. Consequently, we can convert Eq. (\ref{eq2}) into a DDPM-based modeling method according to Eq. (\ref{eq3}):
\begin{equation}
	\begin{array}{l}
		\widehat {Y} = \mathop {\arg \max }\limits_{Y} \log q({Y}) + \log q({X_1},{X_2}...{X_T}|{Y})
		\label{eq3}
	\end{array}
\end{equation}

where ${X_i}$ comprises the LR image $X$ and Gaussian noise added to ${X_{i - 1}}$ at step $i$, while $T$ represents the total diffusion length. In the DDPM, latent variables ${X_1},{X_2}...{X_T}$ are generated by gradually introducing Gaussian noise into the HR image $Y$. Therefore, we can change Eq. (\ref{eq2}) into a modeling method based on DDPM, as shown in Eq. (\ref{eq3}). The architecture of the proposed LDDPM is shown in Fig. \ref{fig_2}. This builds upon the $T$-stage DDPM. Unlike the direct reconstruction of HR images at each iterative step, LDDPM employs a UNet network to predict the noise $\varepsilon$ at the current $i-{th}$ step of ${X_i}$. We incorporate a conditional encoding mechanism into LDDPM, which includes conditional encoding based on an adaptive multi-head attention mechanism and conditional encoding based on a VAE.

In the conditional encoding using the adaptive multi-head attention mechanism, we map the LR image features encoded by the conditional encoder to the middle layer of UNet using a multi-head attention mechanism. This approach guides the UNet network to learn more latent features from the LR images. In contrast, for VAE-based conditional encoding, we sample random feature vectors from the LR image $X$ to serve as conditional features ${F_R}$. These are combined with the mean map ${F_\mu }$ and variance map ${F_\sigma }$ decomposed from the feature vector ${F_X}$ of the UNet encoder, thereby transferring the conditional features to the hidden space. The VAE not only effectively compensates for the information lost when enlarging the LR image but also constrains the solution space for reconstructing the HR image, making it easier for the model to learn the current noise $\varepsilon$.

For the output features ${F_g}$ of the UNet network encoder, we employ a normalized flow to allow the model to better infer more complex probability distribution deviations. During training, to ensure high-quality sampling by the LDDPM model, we employ a GAN to learn the multimodal distribution of ${X_{i - 1}},{X_i}$, replacing the simple Gaussian distribution learned by the original DDPM. This approach reduces the Kullback–Leibler (KL) divergence between the denoising model's noise probability distribution and the actual model's noise probability distribution.

\subsection{Denoising Diffusion Probabilistic Model}
\begin{figure}[H]
	\centering
	\includegraphics[width=\linewidth]{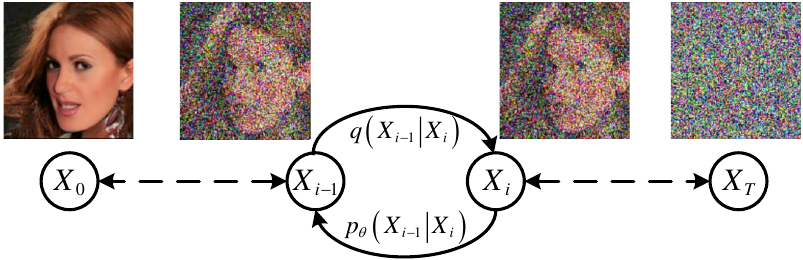}
	\caption{Overview of the forward and inverse diffusion processes of the DDPM. The forward diffusion is from left to right, and the inverse diffusion is from right to left, where $\theta $ denotes the learnable parameters.}
	\label{fig_3}
\end{figure}

In DDPM, we define the HR image $Y$ as the target variable, where $q\left( Y \right)$ represents the data distribution of the target variable. As depicted in Fig. \ref{fig_3}, DDPM consists of forward and reverse diffusion. The aim of DDPM's forward diffusion is to map $Y$ to a multidimensional normal distribution (Gaussian noise) via a Markov chain, as described in Eq. (\ref{eq4}):
\begin{equation}
	\begin{array}{l}
		q\left( {{X_1}, \cdots ,{X_T}\left| {{X_0}} \right.} \right) = \prod\limits_{i = 1}^T q \left( {{X_i}\left| {{X_{i - 1}}} \right.} \right)
		\label{eq4}
	\end{array}
\end{equation}

In this process, we define ${X_0}$ as $Y$ and ${X_i}$ and $Y$ as variables with the same dimension. $T$ represents the number of diffusion steps, and $q\left( {{X_i}\left| {{X_{i - 1}}} \right.} \right)$ is defined as a Gaussian distribution ${\cal N}\left( {{X_i};\sqrt {1 - {\beta i}} {X{i - 1}},{\beta _i}I} \right)$ related to the constant ${\beta _i}$. A small amount of Gaussian noise is added at each diffusion step, eventually transforming the HR image into a multidimensional Gaussian distribution with independent dimensions. The DDPM's reverse diffusion is based on sampling from a Gaussian distribution to generate HR images, as shown in Eq. (\ref{eq5}):
\begin{equation}
	\begin{array}{l}
		{p_\theta }\left( {{X_0}, \cdots ,{X_{T - 1}}\left| {{X_T}} \right.} \right) = \prod\limits_{i = 1}^T {{p_\theta }} \left( {{X_{i - 1}}\left| {{X_i}} \right.} \right),\\
		{p_\theta }\left( {{X_{i - 1}}\left| {{X_i}} \right.} \right){\rm{ = }}{\cal N}\left( {{X_{i - 1}};{\mu _\theta }\left( {{X_i},i} \right),\sigma I} \right),\\
		where{\rm{  }}p\left( {{X_T}} \right) = {\cal N}\left( {0,I} \right)
		\label{eq5}
	\end{array}
\end{equation}

Through Eq. (\ref{eq5}), the model gradually eliminates Gaussian noise, ultimately generating HR images that conform to the target distribution. Notably, during model training, we train only the mean function ${\mu _\theta }$ and set the variance ${\sigma _\theta }$ as a constant, thereby allowing the model to generate samples of HR images. According to the reparameterization, ${\mu _\theta }$ can be rewritten as Eq. (\ref{eq6}):
\begin{equation}
	\begin{array}{l}
		
		{\mu _\theta }\left( {{X_i},i} \right) = \frac{1}{{\sqrt {{a_i}} }}\left( {{X_i} - \frac{{{\beta _i}}}{{\sqrt {1 - {{\bar \alpha }_i}} }}{\varepsilon _\theta }\left( {{X_i},i} \right)} \right),\\
		where \quad {\alpha _i} = 1 - {\beta _i},\overline {{\alpha _i}}  = \prod\limits_{s = 1}^i {{\alpha _s}} 
		
		\label{eq6}
	\end{array}
\end{equation}

Finally, DDPM can be interpreted as predicting the noise $\varepsilon$ added in the $i^{th}$ step from ${X_i}$ given image $Y$, noise $\varepsilon$, and step number $i$. To achieve this, the model must learn effective feature information from ${X_i}$, $\varepsilon$, and step number $i$ to map the HR image $Y$ to the corresponding noise value gradually according to a specified rule and to generate a data distribution close to $Y$ during reverse diffusion based on the noise value. Therefore, the loss function of DDPM is defined by Eq. (\ref{eq7}):

\begin{equation}
	\begin{array}{l}
		
		{L_{DDPM}} = {\mathbb{E}_{i,{X_{_0}},\varepsilon }}\left[ {{{\left\| {\varepsilon  - {\varepsilon _\theta }\left( {\sqrt {{{\bar \alpha }_i}} {X_0} + \sqrt {1 - {{\bar \alpha }_i}} \varepsilon ,i} \right)} \right\|}^2}} \right]
		
		\label{eq7}
	\end{array}
\end{equation}

\subsection{Conditional Encoding Mechanism}
In the SISR task, the purpose of LDDPM is to model the conditional distribution $P(X|Y)$. We can control the reconstruction of HR images by encoding the LR image $X$ as the conditional input to the function ${\varepsilon _\theta }\left( \cdot \right)$. However, if the current noise image ${X_i}$ and $X$ are directly stacked together for conditional sampling within the UNet network, UNet is prone to neglecting perceptually context-dependent detail features and requires expensive function evaluations in the pixel space to better extract noise features. Therefore, encoding the LR image $X$ in DDPM for use as the model's conditional input merits further study because this can enhance the model's learning of the noise distribution.

\textbf{Conditional Encoding Based on Adaptive Multi-Headed Attention Mechanism: }For the modeling condition of LDDPM, we do not only stack the current noise image ${X_i}$ and LR image $X$ together; inspired by Rombach et al. and Qin et al. \cite{DBLP:journals/access/QinYWWLW21}, we also designed a conditional encoding method based on the adaptive multi-head attention mechanism. First, we designed an encoder ${\cal T}_\theta \left( \cdot \right)$ based on the LR image. This encoder projects the LR image $X$ onto the same feature dimension as the middle layer of UNet, and then the multi-head attention mechanism in UNet is used to perform the dot product between the middle layer features and the features projected by ${\cal T}_\theta \left( \cdot \right)$. The calculation method for the conditional adaptive multi-head attention mechanism is given by Eq. (\ref{eq8}):
\begin{equation}
	\begin{array}{l}
		
		\left( {Q,K,V} \right) =  softmax \left( {\frac{{Q{K^T}}}{{\sqrt d }}} \right) \cdot V\\
		where  \quad Q = W_Q^{\left( k \right)} \cdot {\varphi _k}\left( {{X_i}} \right),\\
		\quad  \quad  \quad  \quad K = W_K^{\left( k \right)} \cdot {\tau _\theta }\left( X \right),\\
		\quad  \quad  \quad  \quad V = W_V^{\left( k \right)} \cdot {\tau _\theta }\left( X \right)
		\label{eq8}
	\end{array}
\end{equation}

where ${\varphi k}\left( \cdot \right)$ represents the flattening feature operation, and $W_Q^{\left( k \right)}$, $W_K^{\left( k \right)}$, and $W_V^{\left( k \right)}$ are the projection matrices of the $k$-th intermediate layer of UNet. Notably, in the Glow model based on normalized flows, we replace the mask matrix with the feature matrix output by the conditional encoder ${\cal T}\theta \left( \cdot \right)$, allowing the model to better learn the current noise $\varepsilon$ at the $i$-th step.

\textbf{Conditional Encoding based on Variational Auto-Encoder: }We also designed a CVAE \cite{DBLP:conf/www/LiangKHJ18}, which allows ${X_i}$ to retain more diverse feature information about the condition. CVAE models the variational inference of the latent variable $Z$ and the observed variable $C \in {X_{1,2...T}}$ using the reparameterization trick. CVAE can project ${X_i}$ into the latent space, thereby learning the latent conditional probability distribution. The decoder of UNet learns the feature map ${F_g}$, which is used to predict the noise $\varepsilon$ in UNet by referring to the conditional feature ${F_c}$ output of CVAE. The CVAE model mainly comprises a feature encoder, latent variable $Z$, and feature decoder, as depicted in Figure \ref{fig_2}. The feature encoder of CVAE is used to fit the likelihood function, which is calculated using Eq. (\ref{eq9}):
\begin{equation}
	\begin{array}{l}
		
		{P_\theta }(C\left| Z \right.) = \prod\limits_{i = 1}^T {\cal N} \left( {{X_i};{\mu _\theta }\left( {{Z_i}} \right),{\sigma _\theta }\left( {{Z_i}} \right)I} \right)
		
		\label{eq9}
	\end{array}
\end{equation}

The value of the likelihood function depends on the results of functions ${\mu _\theta }$ and ${\sigma _\theta }$, which learn the mean and variance of the Gaussian distribution, respectively. They learn the relationships between pixels and represent them using a probabilistic model. In CVAE, the latent variable $Z$ can be represented by ${\mu _\theta }$ and ${\sigma _\theta }$, calculated as shown in Eq. (\ref{eq10}).
\begin{equation}
	\begin{array}{l}
		{Z_i} = {\mu _\theta }\left( {{X_i}} \right) + \delta  \cdot {\sigma _\theta }\left( {{X_i}} \right),\\
		where   \quad \delta  \sim {\cal N}\left( {0,I} \right)
		\label{eq10}
	\end{array}
\end{equation}

The latent variable $Z$ is sampled from the Gaussian distribution $Q(Z) = N \sim (0,I)$, as shown in Figure \ref{fig_2}. To ensure that randomness is introduced during sampling and that the probability distribution learned by CVAE is close to the Gaussian distribution, we use KL divergence to optimize CVAE, calculated using Eq. (\ref{eq11}):
\begin{equation}
	\begin{array}{l}
		
		{D_{KL}}\left( {P\left( {Z\left| C \right.} \right)\left\| {Q\left( Z \right)} \right.} \right) = E\left[ {\log P(Z\left| C \right.) - Q(Z)} \right]\\
		{\rm{                                  }} = \frac{1}{2}\left( { - \sum\limits_i {\left( {\log \sigma _i^2 + 1} \right)}  + \sum\limits_i {\sigma _i^2}  + \sum\limits_i {\mu _i^2} } \right)
		\label{eq11}
	\end{array}
\end{equation}

In the forward diffusion of LDDPM, we first replace the Gaussian distribution input to the feature decoder with the latent variable $Z$. We then project the probability distribution output from the feature decoder into the spatial domain using a convolutional layer, thereby obtaining the conditional probability mapping feature ${F_R}$. Finally, to map the conditional probability ${F_R}$ to the output of the UNet encoder, we use a convolutional layer to learn the mean ${F_\mu }$ and variance ${F_\delta }$ of the feature map ${F_X}$ output by the UNet encoder. We fuse the calculation results of the conditional probabilities ${F_R}$ and ${F_X}$ to obtain fused feature ${F_g}$. The mean ${F_\mu }$ and variance ${F_\delta }$ are the spatial variables of the feature mapping, not variables of the Gaussian distribution. Moreover, in the reverse diffusion of LDDPM, we can remove the feature encoder of CVAE using random Gaussian distribution sampling as the input latent variable $Z$ to the feature decoder, thus reducing the computational cost of the model without affecting the image reconstruction effect.

\subsection{Optimized Denoising Diffusion Probabilistic Model}
\textbf{Glow-based Model Optimization: }The objective of this study was to allow the posterior encoder of LDDPM to accurately reconstruct HR images. The better the prior encoder of LDDPM can learn the data distribution, the more effectively the posterior encoder can reconstruct the HR images. Therefore, in LDDPM, we not only utilize the Gaussian distribution in CVAE to parameterize the prior and posterior encoders of UNet, thereby learning the diversity of conditional feature information, but also apply the Glow model, which is based on normalized flows \cite{DBLP:conf/nips/KingmaD18}, to the feature map ${F_g}$, enriching it with more diverse feature information. Glow is a type of flow model. To reduce the computational overhead, we employ a Glow model with three superficial layers. Each superficial layer consists of a squeeze operation and flow step, with each flow step comprising ActNorm, a 1x1 convolutional layer, and a coupling layer. The designed Glow model operates based on the Gaussian distribution output by CVAE. As the noise evolves, the bijective function ${f_\theta }\left( \cdot \right)$ is used to transform simple distributions into more complex ones. This approach allows the posterior encoder to reconstruct the complex data distribution of the HR images more easily. The computation of Glow is expressed by Eq. (\ref{eq12}):
\begin{equation}
	\begin{array}{l}
		{p_\theta }(C\left| {{F_g}} \right.){\rm{ }} = N\left( {{f_\theta }(C);{\mu _\theta }({F_g}),{\sigma _\theta }({F_g})} \right)\left| { det \frac{{\partial {f_\theta }(C)}}{{\partial C}}} \right|
		\label{eq12}
	\end{array}
\end{equation}

\textbf{GAN-based model Optimization: }In DDPM, the KL divergence  ${D_{KL}}(C,Y) = {\mathbb{E}_X}\log \frac{C}{Y}$ quantifies the information loss when the distribution $Y$ is used as a substitute for the distribution $C$. To ensure a higher match degree between the real HR image and reconstructed HR image's data distributions, LDDPM aims to minimize the KL divergence between the data distribution of the denoising model in the reverse diffusion ${p_\theta }\left( {{X_{i - 1}}\left| {{X_i}} \right.} \right)$ and that in the forward diffusion  $q\left( {{X_{i - 1}}\left| {{X_i}} \right.} \right)$. As demonstrated by Xiao et al. \cite{DBLP:conf/iclr/XiaoKV22}, Gaussian noise is added progressively during forward diffusion, and the data distribution approximates a unimodal Gaussian distribution. By contrast, as the step size increases, the data distribution shifts from Gaussian to a more complex distribution. Therefore, we designed a conditional GAN to estimate the real denoising distribution $q\left( {{X_{i - 1}}\left| {{X_i}} \right.} \right)$, enabling LDDPM to possess stronger expressiveness and model multimodal denoising distributions.

\begin{figure}[H]
	\centering
	\includegraphics[width=\linewidth]{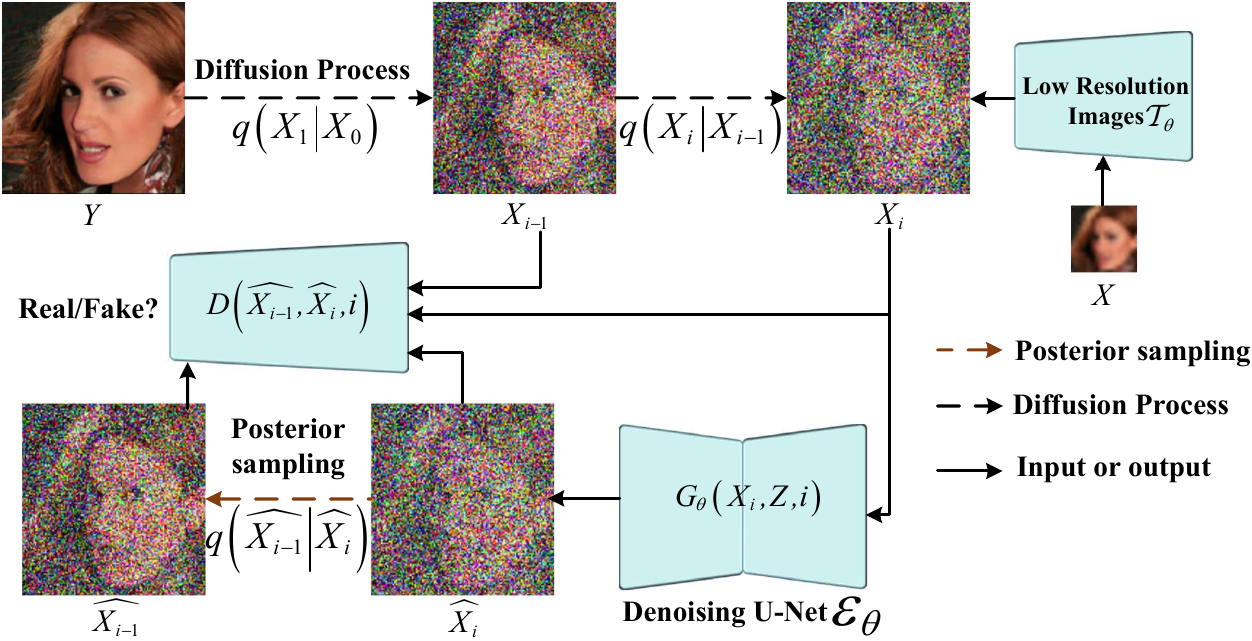}
	\caption{Denoising of LDDPM based on GAN}
	\label{fig_4}
\end{figure}

The goal of our conditional GAN is to minimize the adversarial loss function, thereby minimizing ${D_{KL}}(C,Y)$ and enhancing the matching degree between the probability distributions of reverse diffusion in LDDPM ${p_\theta }\left( {{X_{i - 1}}\left| {{X_i}} \right.} \right)$ and the real denoising distribution in the forward diffusion $q\left( {{X_{i - 1}}\left| {{X_i}} \right.} \right)$. Our proposed conditional GAN is shown in Figure \ref{fig_4}. It features a time-dependent discriminator ${D_\phi }\left( \cdot \right)$, which takes $\widehat {{X_{i - 1}}}$ and $\widehat {{X_i}}$ computed from the noise $\varepsilon $ as input, and it outputs the confidence scores for $\widehat {{X_{i - 1}}}$ and $\widehat {{X_i}}$. The discriminator is trained using Eq. (\ref{eq13}):
\begin{equation}
	\begin{footnotesize}
		\begin{array}{l}
			{{L_{adv}} = {{\min }_\phi }\sum\limits_{i \ge 1} {{\mathbb{E}_{q\left( {{X_i}} \right)}}} \left[ {{\mathbb{E}_{q\left( {\widehat {{X_{i - 1}}}\left| {{{\widehat X}_i}} \right.} \right)}}\left[ {- \log \left( {{D_\phi }\left( {\widehat {{X_{i - 1}}},\widehat {{X_i}},i} \right)} \right)} \right]} \right.}\\
			{\left. { + {\mathbb{E}_{{p_\theta }\left( {\widehat {{X_{i - 1}}}\left| {{{\widehat X}_i}} \right.} \right)}}\left[ { - \log \left( {1 - {D_\phi }\left( {\widehat {{X_{i - 1}}},\widehat {{X_i}},i} \right)} \right)} \right]} \right]}
			
		\end{array}
	\end{footnotesize}
	\label{eq13}
\end{equation}

The output of the GAN generator is the noise distribution $\varepsilon $. Therefore, $\widehat {{X_{i - 1}}}$ and $\widehat {{X_i}}$ can be computed using Eq. (\ref{eq14}):
\begin{equation}
	\begin{array}{l}
		\widehat {{X_{i - 1}}} = \sqrt {{{\bar \alpha }_{i - 1}}} {X_0} + \sqrt {1 - {{\bar \alpha }_{i - 1}}} {\varepsilon _\theta },\\
		\widehat {{X_i}} = \sqrt {{\alpha _{i}}} \widehat {{X_{i - 1}}} + \sqrt {1 - {\alpha _{i}}} \varepsilon, \\
		where   \quad \varepsilon  \sim {\cal N}\left( {0,I} \right)
		\label{eq14}
	\end{array}
\end{equation}

Compared with DDPM, the current LDDPM reconstructs HR images with more complex distributions during reverse diffusion, and LDDPM is an implicit model. The forward diffusion of LDDPM remains a Gaussian noise addition; hence, regardless of the step size or complexity of the data distribution, the forward diffusion $q\left( {{X_{i - 1}}\left| {{X_i}} \right.} \right)$ retains its Gaussian property. Consequently, the reverse diffusion of LDDPM ${p_\theta }\left( {{X_{i - 1}}\left| {{X_i}} \right.} \right)$ can be represented by Eq. (\ref{eq15}):
\begin{equation}
	\begin{array}{l}
		{p_\theta }\left( {{X_{i - 1}}\left| {{X_i}} \right.} \right) = \int {{p_\theta }} \left( {{\varepsilon _\theta }\left| {{X_i}} \right.} \right)q\left( {\widehat {{X_{i - 1}}}\left| {\widehat {{X_i}}} \right.,{\varepsilon _\theta }} \right)d{\varepsilon _\theta }\\
		{\rm{   }} = \int p (Z)q\left( {\widehat {{X_{i - 1}}}\left| {\widehat {{X_i}}} \right.,{\varepsilon _\theta } = {G_\theta }\left( {{X_i},Z,i} \right)} \right)dZ
		\label{eq15}
	\end{array}
\end{equation}

Here, ${p_\theta }\left( {{\varepsilon _\theta }\left| {{X_i}} \right.} \right)$ is the implicit distribution introduced by the GAN's generator ${G_\theta }\left(  \cdot  \right)$, which takes as input ${X_i}$, $Z \sim p(Z) = {\cal N}(Z;0,I)$, and the diffusion step number $i$.
\subsection{Loss Function for Training}
During training, LDDPM gradually maps $Y$ to a Gaussian distribution via a Markov chain. However, the noise error increases with an increase in the number of iterations. To reduce the perceptual distance between the reconstructed HR image $\widehat Y$ and real HR image $Y$ caused by noise errors during training, we use content- and style-aware guidance to assist LDDPM in reconstructing the HR image. Our idea is that the real denoising image ${X_i}$ at each iteration step must retain the content information and pass the style information to the denoising image $\widehat {{X_i}}$ reconstructed by LDDPM.

\textbf{Content Loss: }Ledig et al. \cite{DBLP:conf/cvpr/LedigTHCCAATTWS17} demonstrated that the mean square error (MSE) loss function tends to lack high-frequency detail information, which can lead to overly smooth textures in images reconstructed in SISR tasks. Hence, we compute the loss value of images ${X_i}$ and ${\widehat X_i}$ based on the VGG-19 content loss (CL), thereby preserving more detail features of image ${X_i}$. Unlike Ledig et al., we also calculate the loss value of the real HR image $Y$ and the reconstructed HR image $\widehat Y$ for the differences between the pixels. The content loss calculation for LDDPM is shown in Eq. (\ref{eq16}):
\begin{equation}
	\begin{array}{l}
		{L_{content}} = \frac{1}{{{W_{i,j}}{H_{i,j}}}}\sum\limits_{x = 1}^{{W_{i,j}}} {\sum\limits_{y = 1}^{{H_{i,j}}} {\left( {vg{g_{i,j}}{{\left( {{X_i}} \right)}_{x,y}}} \right.} } \\
		- {\left. {vg{g_{i,j}}{{\left( {{G_{{\theta _G}}}\left( {{{\widehat X}_i}} \right)} \right)}_{x,y}}} \right)^2}{\rm{ + }}\left\| {Y} \right. - {\left. {\widehat {Y}} \right\|^1}\\
		where \quad \widehat {Y} = ({X_i} - \sqrt {1 - {{\bar \alpha }_i}} {\varepsilon _\theta }) \times \frac{1}{{\sqrt {{{\bar \alpha }_i}} }}
		\label{eq16}
	\end{array}
\end{equation}

\textbf{Style Loss: }Park et al. \cite{DBLP:conf/cvpr/ParkL19} used VGG-19 (Relu1\_2, Relu2\_2, Relu3\_4, and Relu4\_1) to extract feature maps and calculated the mean and variance of the feature maps to reduce the style loss (SL) of the feature maps. LDDPM refers to the method of Park et al. to calculate the style loss value of ${X_i}$ and $\widehat {{X_i}}$, and the calculation is shown in Eq. (\ref{eq17}):
\begin{equation}
	\begin{footnotesize}
		\begin{array}{l}
			
			{L_{{\rm{style}}}} = \frac{1}{{{W_{i,j}}{H_{i,j}}}}\sum\limits_{x = 1}^{{W_{i,j}}} {\sum\limits_{y = 1}^{{H_{i,j}}} {\left( {{\mu _\theta }\left( {vg{g_{i,j}}{{\left( {{X_i}} \right)}_{x,y}}} \right)} \right.} } \\
			{\rm{        }} - {\left. {{\mu _\theta }\left( {vg{g_{i,j}}{{\left( {{G_{{\theta _G}}}\left( {{{\widehat X}_i}} \right)} \right)}_{x,y}}} \right)} \right)^2}\\
			{\rm{        }} + {\left( {{\sigma _\theta }\left( {vg{g_{i,j}}{{\left( {{X_i}} \right)}_{x,y}}} \right) - \left. {{\sigma _\theta }\left( {vg{g_{i,j}}{{\left( {{G_{{\theta _G}}}\left( {{{\widehat X}_i}} \right)} \right)}_{x,y}}} \right)} \right)} \right)^2}\\
			{\rm{       }}
			
			\label{eq17}
		\end{array}
	\end{footnotesize}
\end{equation}

The total loss function of the LDDPM model is shown by Eq.  (\ref{eq18}):
\begin{equation}
	\begin{array}{l}
		{L_{total = }}{L_{DDPM}} + {D_{KL}}\left( {P\left( {Z\left| C \right.} \right)\left\| {Q\left( Z \right)} \right.} \right)\\
		{\rm{       }} + {L_{adv}} + {L_{content}} + {L_{{\rm{style}}}}

		\label{eq18}
	\end{array}
\end{equation}

\section{Experiments and Analysis}
In this section, the experimental details of the model are presented, and the applicability of the proposed model to SISR tasks is demonstrated. First, the experimental setup of the model is described in detail. Experimental results of the proposed model and other advanced models are then presented. Finally, the network structure of the model is verified by ablation experiments, and we prove that the network structure designed in this study can recover image details and texture features well and reconstruct realistic HR images in the SISR task.
\subsection{Experimental Settings}
\textbf{Datasets:} 
We evaluated the adaptability of LDDPM across a range of datasets. This included face datasets with an upscaling factor of 8× and general-purpose datasets with scaling factors of 2×, 3×, 4×, 8×, 16×, and 32×. For the face datasets, we employed the Flickr-Faces-High-Quality (FFHQ) (Karras et al. \cite{DBLP:journals/pami/KarrasLA21}, 2019) and CelebFaces Attribute (CelebA) (Liu et al.\cite{liu2018large}, 2018) datasets.

FFHQ is a high-quality face dataset containing 70K images that represent diverse attributes, such as age, ethnicity, image background, and facial features. For our experiments, we selected 52K images from FFHQ for training. CelebA, which is rich in facial attribute data, encompasses 200K images with a wide range of poses and complex backgrounds. For the CelebA dataset, the settings of SrDiff [Li et al., 2022] \cite{DBLP:journals/ijon/LiYCCFXLC22} were followed, and 5,000 images were selected for testing, with the remaining images used for training. The FFHQ dataset primarily served as the pretraining dataset for LDDPM. For SISR tasks with a more general application, we used the DIV2K (Agustsson et al. \cite{DBLP:conf/cvpr/AgustssonT17}, 2017), Flickr2K (Lim et al. \cite{DBLP:conf/cvpr/LimSKNL17}, 2017), and CIFAR-10 datasets. We selected 800 images from DIV2K and 2650 images from Flickr2K for training and set aside 100 images from DIV2K for testing. To evaluate the versatility of our model, we chose Set5, Set14, Urban100, General100, and Manga109 as the test sets.

Moreover, to enhance the resilience of the model, we implemented the image degradation algorithm proposed by Zhang et al. \cite{DBLP:conf/iccv/0008LGT21} on the LR images. This included random combinations of blur, down-sampling, and noise degradation. Blur degradation was simulated using isotropic and anisotropic Gaussian blurring. Down-sampling degradation was achieved by randomly selecting from the nearest-neighbor, bilinear, and cubic spline interpolation methods. Noise degradation involved adding various levels of Gaussian noise, JPEG compression of different qualities, and sensor noise generated by reversing image signal processing (ISP) to the LR images.

\textbf{Experimental Parameters: }The models in this study were trained using eight GeForce RTX 3090 24 GB graphics cards, four TITAN V 16 GB cards, and one A100 80 GB card. Table \ref{tab:table1} lists the model parameter settings used for training and testing on CelebA, DIV2K, CIFAR-10, and the combined DIV2K and Flickr2K (DIFL2K) datasets. These settings were consistently used to produce the primary results presented in the tables in this paper. The same configurations were also used for all LDDPM variations in the ablation experiments.

We evaluated performance by employing a diverse set of metrics: peak signal-to-noise ratio (PSNR) gauged pixel-level discrepancies, structural similarity (SSIM) tracked structural variations, and LR-PSNR ensured consistency between the super-resolution output and original LR images. Furthermore, we adopted learned perceptual image patch similarity (LPIPS) and deep image structure and texture similarity (DISTS), both of which prioritize perceptual quality and deep feature likeness. Finally, we employed Fréchet inception distance (FID) as a tool to quantify the statistical discrepancies between LR and HR images, where lower FID values indicate superior quality. We also utilized the inception score (IS) to gauge the diversity and realism of the generated images.
\begin{table}[!ht]
	\caption{\label{tab:table1}Training parameter settings of the model}
	\centering
	\resizebox{.95\columnwidth}{!}{
		\setlength{\tabcolsep}{0mm}{
			\begin{tabular}{ccc}
				\toprule
				\multirow{2}[1]{*}{Training Config} & \multicolumn{1}{c}{\multirow{2}[1]{*}{DIV2K/DIFL2K}} & \multicolumn{1}{c}{\multirow{2}[1]{*}{CelebA/CIFAR-10}} \\
				\multicolumn{1}{c}{} &       &  \\
				\midrule
				
				Inner Channel & 64    & 64 \\
				Channel Multiplier & \multicolumn{1}{c}{(1,2,4,8,8)} & \multicolumn{1}{c}{(1,2,4,8,8)} \\
				Dropout & 0.2   & 0.2 \\
				Diffusion steps  & 1000  & 1000 \\
				Base Learning Rate & 3E-05 & 1E-04 \\
				Learning Rate Schedule & \multicolumn{1}{c}{Cosine Decay} & \multicolumn{1}{c}{Cosine Decay} \\
				Batch Size & 16    & 20 \\
				Training Epochs & 50    & 50 \\
				Exponential Moving Average (EMA) & 0.9999 & 0.9999 \\
				Optimizer & \multicolumn{1}{c}{Adamw \cite{DBLP:conf/iclr/LoshchilovH19}} & \multicolumn{1}{c}{Adamw} \\
				OptimizeR Momentum & ${\beta _1},{\beta _2} = 0.9,0.999$     & ${\beta _1},{\beta _2} = 0.9,0.999$ \\
				Layer Scale & 1E-06 & 1E-06 \\
				\bottomrule
			\end{tabular}%

	}}
\end{table}

\subsection{Comparison of General Image Super-resolution reconstruction Experiments}
In this section, we evaluate the reconstruction performance of LDDPM on common image super-resolution datasets by comparing it with advanced SISR reconstruction models.

First, for standard image super-resolution datasets (×2, ×3, ×4, LR image size of 128), we compared LDDPM with EDSR (Lim et al.,2017), RCAN (Zhang et al.,2018), SAN (Dai et al., 2019), IGNN (Zhou et al.,2020), HAN (Niu et al., 2020), and NLSA (Mei et al., 2021) on the DIV2K dataset. Second, on the DIFL2K dataset, we compared LDDPM with SwinIR (Liang et al., 2021), SwinFIR (Zhang et al., 2022), and EDT (Li et al., 2022). Finally, we compared LDDPM with SR3 (Saharia et al., 2022) after using the CelebA dataset for pre-training and fine-tuning the model parameters on the DIFL2K dataset. Among these models, EDSR, RCAN, SAN, IGNN, HAN, and NLSA primarily use regression-based models. SwinIR, EDT, and SwinFIR primarily use GAN-based models. SR3 primarily uses DDPM-based models. Table \ref{tab:table2} presents the experimental results of these classic SISR reconstruction models.

Compared with other advanced models, our LDDPM showed superior performance in reconstructing HR images on multiple test sets, particularly on Urban100, thus validating the effectiveness of LDDPM in reconstructing high-frequency HR images and offering a new perspective for SISR tasks. Additionally, Table \ref{tab:table2} shows that when we used CelebA as the pre-training dataset for LDDPM, improvement is noticeable in the reconstruction performance of LDDPM, indicating that pre-training models can impose certain constraints on the solution space of HR images.

\begin{figure*}[ht]
	\centering
	\includegraphics[width=\textwidth]{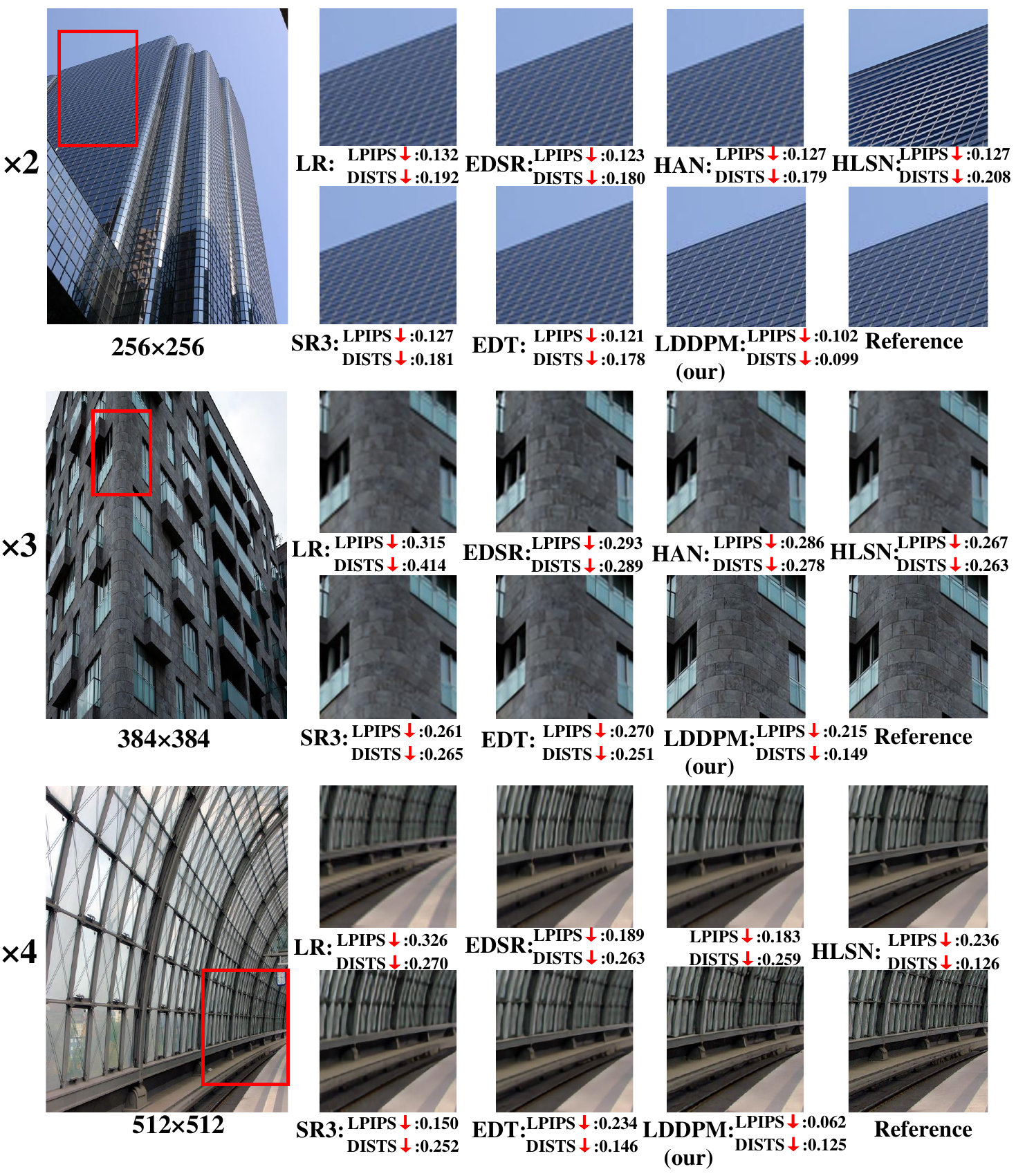}
	\caption{Visualization of the results by different models on the Urban100 dataset (×2, ×3, ×4, LR image size of 128).}
	\label{fig_5}
\end{figure*}

We visualized the HR images reconstructed using advanced models, as shown in Figure \ref{fig_5}. Figure \ref{fig_5} shows the reconstruction of complex image details and textures using the current advanced improvement. However, LDDPM not only effectively constrains the solution space of HR images but also excels in reconstructing HR images with high-frequency details.

To verify that LDDPM can effectively reconstruct lower-resolution images across different resolution datasets, we re-evaluated the image reconstruction capability of the model on the DIFL2K dataset (×4, LR image size of 64). Table \ref{tab:table2-2} compares LDDPM with GAN-based models (including SFTGAN, SRGAN, ESRGAN, USRGAN, SPSR, and BebyGAN) and DDPM-based models (namely, LDM and DiffIR on the DIV2K dataset) using LPIPS and DISTS as evaluation metrics. Table \ref{tab:table2-2} shows that LDDPM achieved the lowest LPIPS scores across multiple test sets. Additionally, compared with the advanced DiffIR model, LDDPM achieved lower DISTS scores on the General100 and Urban100 datasets.

\begin{figure*}[ht]
	\centering
	\includegraphics[width=\textwidth]{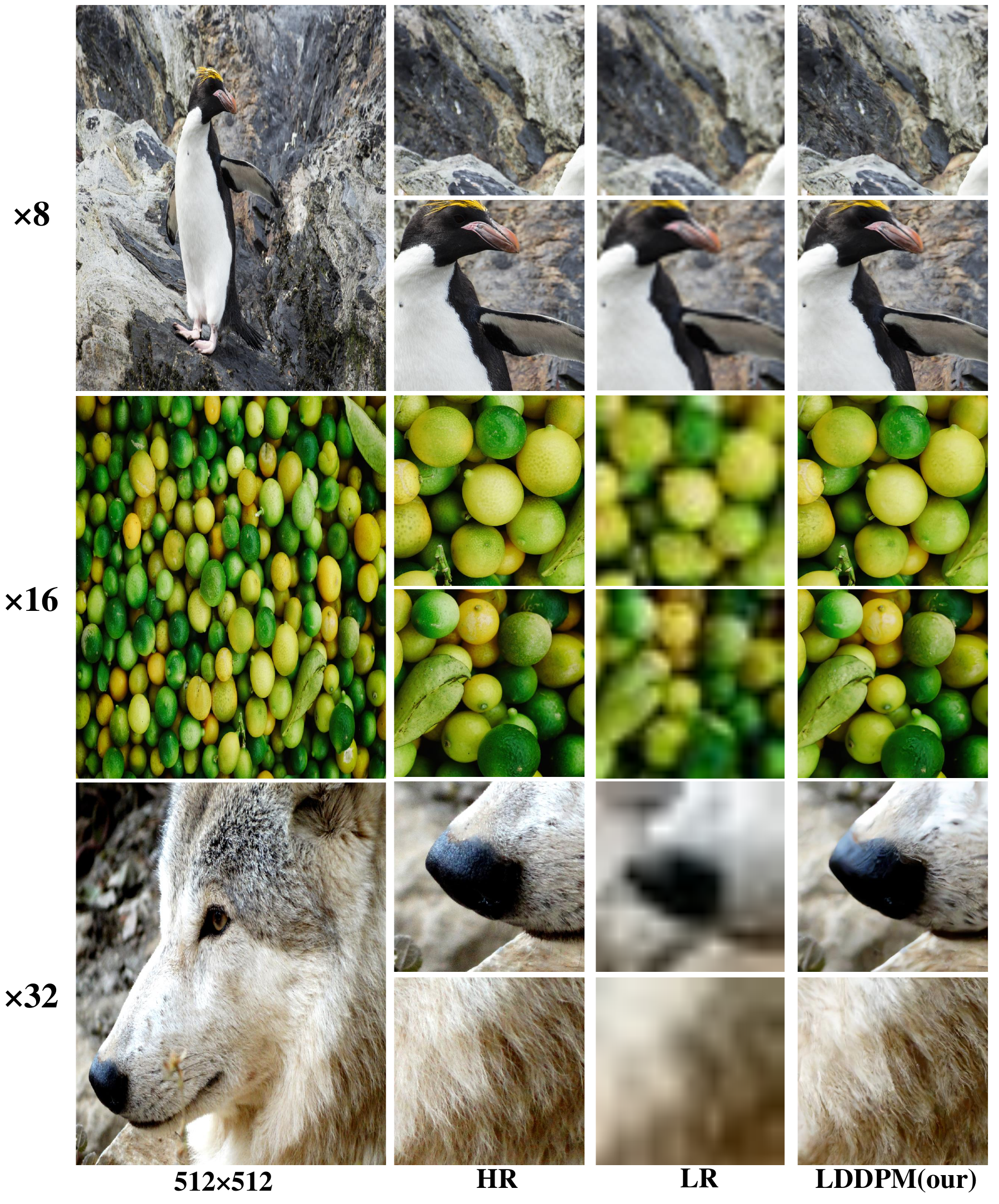}
	\caption{Results of LDDPM on the DIFL2K dataset (×8, ×16, ×32,  LR image sizes of  64, 32, and 16, respectively) are visualized.}
	\label{fig_12}
\end{figure*}

To demonstrate that LDDPM can effectively reconstruct HR images even at higher resolutions, we visualized the reconstructed HR images of the model at different resolutions on the DIFL2K dataset (×8, ×16, ×32, LR image sizes of  64, 32, and 16). The visualization results are shown in Figure \ref{fig_12}. Evidently, LDDPM can reconstruct highly detailed images across a variety of resolutions. This includes the texture of stones, penguin hair, wolf fur and nose, and fruit. Therefore, LDDPM is capable of reconstructing HR images with good perceptual quality from LR images.

Table \ref{tab:table2-3} presents a quantitative comparison of LDDPM and other state-of-the-art generative models on the CIFAR-10 dataset (×2, LR image size of 16) using the FID and IS metrics. Table \ref{tab:table2-3} shows that the FID for the images reconstructed by our proposed LDDPM model reached 2.98, outperforming other advanced models. LDDPM can reconstruct high-quality images. In addition, with an IS of 9.62, the images reconstructed by LDDPM maintained a good level of sample diversity.
\begin{table*}[]
	\caption{\label{tab:table2}Quantitative comparison of LDDPM models with advanced models on classical image super-resolution reconstruction data (×2, ×3, ×4, LR image size of 128)}
	\centering
	\resizebox{2\columnwidth}{!}{
		\setlength{\tabcolsep}{1mm}{
			\begin{tabular}{ccccccccccc}
				\toprule
				\multicolumn{1}{c}{} & \multicolumn{1}{c}{} & \multirow{2}[4]{*}{\textbf{Training Dataset}} & \multicolumn{2}{c}{\textbf{Set5}} & \multicolumn{2}{c}{\textbf{Set14}} & \multicolumn{2}{c}{\textbf{Urban100}} & \multicolumn{2}{c}{\textbf{Manga109}} \\
				\textbf{Models} &  \multicolumn{1}{c}{\textbf{Scale}} & \multicolumn{1}{c}{} & \multicolumn{1}{c}{PSNR$\uparrow$(dB)} & \multicolumn{1}{c}{SSIM$\uparrow$} & \multicolumn{1}{c}{PSNR$\uparrow$(dB)} & \multicolumn{1}{c}{SSIM$\uparrow$} & \multicolumn{1}{c}{PSNR$\uparrow$(dB)} & \multicolumn{1}{c}{SSIM$\uparrow$} & \multicolumn{1}{c}{PSNR$\uparrow$(dB)} & \multicolumn{1}{c}{SSIM$\uparrow$} \\
				\midrule
				EDSR (CVPR,2017) \cite{DBLP:conf/cvpr/LimSKNL17} & ×2    & DIV2K & 38.11 & 0.9602 & 33.92 & 0.9195 & 32.93 & 0.9351 & 39.10  & 0.9773 \\
				
				RCAN (ECCV,2018) \cite{DBLP:conf/eccv/ZhangLLWZF18} & ×2    & DIV2K & 38.27 & 0.9614 & 34.12 & 0.9216 & 33.34 & 0.9384 & 39.44 & 0.9786 \\
				
				SAN (CVPR,2019) \cite{DBLP:conf/cvpr/DaiCZXZ19} & ×2    & DIV2K & 38.31 & 0.9620  & 34.07 & 0.9213 & 33.10  & 0.9370  & 39.32 & 0.9792 \\
				
				IGNN (NIPS,2020) \cite{DBLP:conf/nips/ZhouZZL20} & ×2    & DIV2K & 38.24 & 0.9613 & 34.07 & 0.9217 & 33.23 & 0.9383 & 39.35 & 0.9786 \\
				
				HAN (ECCV,2020) \cite{DBLP:conf/eccv/NiuWRZYWZCS20} & ×2    & DIV2K & 38.27 & 0.9614 & 34.16 & 0.9217 & 33.35 & 0.9385 & 39.46 & 0.9785 \\
				
				NLSN (CVPR,2021) \cite{DBLP:conf/cvpr/MeiFZ21} & ×2    & DIV2K & 38.34 & 0.9618 & 34.08 & 0.9231 & 33.42 & 0.9394 & 39.59 & 0.9789 \\
				
				LDDPM & ×2    & DIV2K & 36.75 & 0.9408 & 32.89 & 0.9142 & 31.30  & 0.9202 & 36.21 & 0.9636 \\
				
				SwinIR (CVPR,2021) \cite{DBLP:conf/iccvw/LiangCSZGT21} & ×2    & DIFL2K & 38.42 & 0.9623 & 34.46 & 0.9250  & 33.81 & 0.9427 & 39.92 & 0.9797 \\
				
				SwinFIR (arXiv,2022) \cite{DBLP:journals/corr/abs-2208-11247} & ×2    & DIFL2K & 38.57 & 0.9630  & 34.66 & 0.9263 & 34.30  & 0.9459 & 40.30  & 0.9809 \\
				
				EDT (arXiv,2022) \cite{DBLP:journals/corr/abs-2112-10175} & ×2    & DIFL2K & 38.63 & 0.9632 & 34.80  & 0.9273 & 34.27 & 0.9456 & 40.37 & 0.9811 \\
				
				LDDPM & ×2    & DIFL2K & 37.17 & 0.9610  & \textbf{36.87} & 0.9468 & 34.44 & 0.9602 & 41.48 & \textbf{0.9721} \\
				
				SR3 (T-PAMI,2022) & ×2    & CelebA & 32.89 & 0.9578 & 36.55 & 0.9280  & 34.06 & 0.9594 & 36.39 & 0.9525 \\
				
				LDDPM & ×2    & CelebA & \textbf{38.91} & \textbf{0.9639} & 36.27 & \textbf{0.9525} & \textbf{35.14} & \textbf{0.9610} & \textbf{42.96} & 0.9700 \\
				\midrule
				EDSR (CVPR,2017) & ×3    & DIV2K & 34.65 & 0.9280  & 30.52 & 0.8462 & 28.8  & 0.8653 & 34.17 & 0.9476 \\
				
				RCAN (ECCV,2018) & ×3    & DIV2K & 34.74 & 0.9299 & 30.65 & 0.8482 & 29.09 & 0.8702 & 34.44 & 0.9499 \\
				
				SAN (CVPR,2019) & ×3    & DIV2K & 34.75 & 0.9300    & 30.59 & 0.8476 & 28.93 & 0.8671 & 34.30  & 0.9494 \\
				
				IGNN (NIPS,2020) & ×3    & DIV2K & 34.72 & 0.9298 & 30.66 & 0.8484 & 29.03 & 0.8689 & 34.39 & 0.9496 \\
				
				HAN (ECCV,2020) & ×3    & DIV2K & 34.75 & 0.9299 & 30.67 & 0.8483 & 29.10  & 0.8705 & 34.48 & 0.9500 \\
				
				NLSN (CVPR,2021) & ×3    & DIV2K & 34.85 & 0.9306 & 30.70  & 0.8485 & 29.25 & 0.8726 & 34.57 & 0.9508 \\
				
				LDDPM & ×3    & DIV2K & 34.92 & 0.9328 & 31.66 & 0.8495 & \textbf{30.99} & 0.8837 & 34.63 & 0.9519 \\
				
				SwinIR (CVPR,2021) & ×3    & DIFL2K & 34.97 & 0.9318 & 30.93 & 0.8534 & 29.75 & 0.8826 & 35.12 & 0.9537 \\
				
				SwinFIR (arXiv,2022) & ×3    & DIFL2K & 35.13 & 0.9328 & 31.13 & 0.8556 & 30.20  & 0.8885 & 35.53 & 0.9554 \\
				
				EDT (arXiv,2022) & ×3    & DIFL2K & 34.97 & 0.9316 & 30.89 & 0.8527 & 29.72 & 0.8814 & 35.13 & 0.9534 \\
				
				LDDPM & ×3    & DIFL2K & 35.49 & 0.9335 & 31.20  & 0.8635 & 31.23 & 0.8797 & 35.76 & 0.9556 \\
				
				SR3 (T-PAMI,2022) & ×3    & CelebA & 33.32 & 0.9306 & 30.57 & 0.8399 & 28.76 & 0.8574 & 31.14 & 0.9092 \\
				
				LDDPM & ×3    & CelebA & \textbf{36.46} & \textbf{0.9399} & \textbf{31.87} & \textbf{0.8959} & 30.30  & \textbf{0.9245} & \textbf{36.23} & \textbf{0.9589} \\
				\midrule
				EDSR (CVPR,2017) & ×4    & DIV2K & 32.46 & 0.8968 & 28.80  & 0.7876 & 26.64 & 0.8033 & 31.02 & 0.9148 \\
				
				RCAN (ECCV,2018) & ×4    & DIV2K & 32.63 & 0.9002 & 28.87 & 0.7889 & 26.82 & 0.8087 & 31.22 & 0.9173 \\
				
				SAN (CVPR,2019) & ×4    & DIV2K & 32.64 & 0.9003 & 28.92 & 0.7888 & 26.79 & 0.8068 & 31.18 & 0.9169 \\
				
				IGNN (NIPS,2020) & ×4    & DIV2K & 32.57 & 0.8998 & 28.85 & 0.7891 & 26.84 & 0.8090  & 31.28 & 0.9182 \\
				
				HAN (ECCV,2020) & ×4    & DIV2K & 32.64 & 0.9002 & 28.90  & 0.7890  & 26.85 & 0.8094 & 31.42 & 0.9177 \\
				
				NLSN (CVPR,2021) & ×4    & DIV2K & 32.59 & 0.9000    & 28.87 & 0.7891 & 26.96 & 0.8109 & 31.27 & 0.9184 \\
				
				LDDPM & ×4    & DIV2K & 33.14 & 0.9009 & 28.72 & 0.7832 & 26.14 & 0.8032 & 28.72 & 0.8732 \\
				
				SwinIR (CVPR,2021) & ×4    & DIFL2K & 32.92 & 0.9044 & 29.09 & 0.7950  & 27.45 & 0.8254 & 32.03 & 0.9260 \\
				
				SwinFIR (arXiv,2022) & ×4    & DIFL2K & 33.08 & 0.9048 & 29.21 & 0.7971 & 27.87 & 0.8348 & \textbf{32.52} & \textbf{0.9292} \\
				
				EDT (arXiv,2022) & ×4    & DIFL2K & 32.82 & 0.9031 & 29.09 & 0.7939 & 27.46 & 0.8246 & 32.05 & 0.9254 \\
				
				LDDPM & ×4    & DIFL2K & 33.53 & 0.9046 & 29.36 & 0.7932 & 28.32 & 0.8252 & 30.41 & 0.8975 \\
				
				SR3 (T-PAMI,2022) & ×4    & CelebA & 31.30  & 0.8997 & 28.98 & 0.7541 & 26.98 & 0.8306 & 28.72 & 0.8692 \\
				
				LDDPM & ×4    & CelebA & \textbf{33.93} & \textbf{0.9082} & \textbf{29.52} & \textbf{0.7953} & \textbf{28.62} & \textbf{0.8493} & 31.09 & 0.9054 \\
				\bottomrule
			\end{tabular}%
			
		}
	}
\end{table*}

\begin{table*}[h]
	\caption{\label{tab:table2-2}Quantitative comparison of LDDPM with state-of-the-art models on classical image super-resolution data (×4, LR image size of 64)}
	\centering
	\resizebox{1.7\columnwidth}{!}{
		\setlength{\tabcolsep}{1mm}{
			\begin{tabular}{c|ccccccccccc}
				\toprule
				\multicolumn{1}{c}{\multirow{2}[4]{*}{\textbf{Methods}}} & \multirow{2}[4]{*}{\textbf{Models}} & \multicolumn{2}{c}{\textbf{DIV2K100}} & \multicolumn{2}{c}{\textbf{General100}} & \multicolumn{2}{c}{\textbf{Set5}} & \multicolumn{2}{c}{\textbf{Set14}} & \multicolumn{2}{c}{\textbf{Utban100}} \\
				\cmidrule{3-12}    \multicolumn{1}{c}{} & \multicolumn{1}{c}{} & \multicolumn{1}{c}{LPIPS$\downarrow$} & \multicolumn{1}{c}{DISTS$\downarrow$} & \multicolumn{1}{c}{LPIPS$\downarrow$} & \multicolumn{1}{c}{DISTS$\downarrow$} & \multicolumn{1}{c}{LPIPS$\downarrow$} & \multicolumn{1}{c}{DISTS$\downarrow$} & \multicolumn{1}{c}{LPIPS$\downarrow$} & \multicolumn{1}{c}{DISTS$\downarrow$} & \multicolumn{1}{c}{LPIPS$\downarrow$} & \multicolumn{1}{c}{DISTS$\downarrow$} \\
				\midrule
				\multicolumn{1}{c|}{\multirow{6}[2]{*}{GAN-based Methods}} & SFTGAN (CVPR,2018) \cite{DBLP:conf/cvpr/WangYDL18} & 0.133 & 0.073 & 0.094 & 0.099 & 0.080  & 0.108 & 0.131 & 0.113 & 0.134 & 0.106 \\
				& SRGAN (CVPR,2017) \cite{DBLP:conf/cvpr/LedigTHCCAATTWS17} & 0.125 & 0.066 & 0.096 & 0.098 & 0.075 & 0.100   & 0.132 & 0.106 & 0.143 & 0.108 \\
				& ESRGAN (ECCVW,2018) \cite{DBLP:conf/eccv/WangYWGLDQL18} & 0.115 & 0.059 & 0.087 & 0.087 & 0.075 & 0.094 & 0.124 & 0.095 & 0.122 & 0.088 \\
				& USRGAN (CVPR,2020) \cite{DBLP:conf/cvpr/ZhangGT20} & 0.132 & 0.064 & 0.093 & 0.093 & 0.079 & 0.104 & 0.134 & 0.099 & 0.133 & 0.097 \\
				& SPSR (CVPR,2020) \cite{DBLP:conf/cvpr/MaRCCL020} & 0.109 & 0.054 & 0.086 & 0.088 & 0.064 & 0.092 & 0.120  & 0.092 & 0.118 & 0.084 \\
				& BebyGAN (AAAI,2022) \cite{DBLP:conf/aaai/LiZQLL22} & 0.102 & 0.049 & 0.077 & 0.079 & 0.062 & 0.090  & 0.115 & 0.086 & 0.109 & 0.079 \\
				\midrule
				\multicolumn{1}{c|}{\multirow{3}[1]{*}{DDPM-based Methods}} & LDM (CVPR,2022) & 0.196 & 0.096 & 0.165 & 0.142 & 0.148 & 0.153 & 0.203 & 0.135 & 0.181 & 0.128 \\
				& DiffIR (arXiv,2023) \cite{DBLP:journals/corr/abs-2303-09472} & 0.100   & \textbf{0.049} & 0.065 & 0.064 & 0.062 & \textbf{0.074} & 0.102 & \textbf{0.077} & 0.090  & 0.073 \\
				& LDDPM & \textbf{0.099} & 0.067 & \textbf{0.052} & \textbf{0.049} & \textbf{0.062} & 0.086 & \textbf{0.092} & 0.108 & \textbf{0.089} & \textbf{0.062} \\
				\bottomrule
			\end{tabular}%

	}}
	
\end{table*}

\begin{table}[H]
	\caption{\label{tab:table2-3}Quantitative comparison of LDDPM with state-of-the-art models performed on the CIFAR-10 dataset (×2, LR image size of 16). FID and IS were obtained from calculations on 50K samples.}
	\centering
	\resizebox{1\columnwidth}{!}{
		\setlength{\tabcolsep}{0mm}{
			\begin{tabular}{c|ccc}
				\toprule
				\multicolumn{1}{c|}{\textbf{Methods}} & \multicolumn{1}{c}{\textbf{Models}} & \multicolumn{1}{c}{FID@50k$\downarrow$} & \multicolumn{1}{c}{IS$\uparrow$} \\ 	\midrule
				& VAEBM (ICLR,2022) \cite{DBLP:conf/iclr/XiaoKV22} & 12.19 & 8.43 \\
				& DC-VAE (CVPR,2021) \cite{DBLP:conf/cvpr/ParmarLLT21} & 17.90  & 8.20 \\
				\multicolumn{1}{c|}{VAE-based Methods} & D2C (NIPS,2021) \cite{DBLP:conf/nips/SinhaSME21} & 10.15 & \multicolumn{1}{c}{-} \\
				& NVAE (NIPS,2021) \cite{DBLP:conf/nips/VahdatK20} & 51.67 & 5.51 \\ \midrule
				\multicolumn{1}{c|}{Flow-based Methods} & SRFlow (ECCV,2020) \cite{DBLP:conf/eccv/LugmayrDGT20} & 16.89 & 8.42 \\ \midrule
				& AutoGAN (AAAI,2020) \cite{DBLP:conf/aaai/Cao0WGS20} & 5.29  & 8.55 \\
				\multicolumn{1}{c|}{GAN-based Methods} & BigGAN (ICLR, 2019) \cite{DBLP:conf/iclr/BrockDS19} & 14.73 & 9.22 \\
				& StyleGAN2 (CVPR,2019) \cite{DBLP:journals/pami/KarrasLA21} & 8.32  & 9.21 \\
				& SNGAN+ (ICLR,2018) \cite{DBLP:conf/iclr/MiyatoKKY18} & 15.42 & 8.22 \\
				& GLEAN (CVPR,2022) \cite{DBLP:conf/cvpr/ChanWXGL21} & 13.78 & 8.34 \\ \midrule
				& NCSN (NIPS,2019) \cite{DBLP:conf/nips/SongE19} & 25.32 & 8.87 \\
				\multicolumn{1}{c|}{DDPM-based Methods} & NCSNV2 (NIPS,2020) \cite{DBLP:conf/nips/0011E20} & 10.87 & 8.40 \\
				& LDM (CVPR,2022) & 3.86  & 9.57 \\
				& LDDPM & \textbf{2.98} & \textbf{9.62} \\\midrule
			\end{tabular}%

		}
	}
\end{table}

\subsection{Comparison of Experimental Results of Super-resolution Reconstruction of Face Images}

In this section, we compare the LDDPM, ESRGAN (Wang et al., 2018), ProgFSR (Kim et al., 2019), SRFlow, SRDiff (Li et al., 2022), GLEAN (Kelvin et al., 2021), LDM (Rombach et al., 2022), and SR3 models on the CelebA dataset, as shown in Table \ref{tab:table3}. LDM, SRDiff, and SR3 are DDPM-based models, while ProgFSR, ESRGAN, and GLEAN are GAN-based models, and SRFlow is based on normalized flow. In accordance with Table \ref{tab:table3}, LDDPM outperformed all other models in terms of the evaluation metrics. Compared to the advanced LDM model, LDDPM improves the PSNR, SSIM, and LR-PSNR of reconstructed images by 0.62, 0.042, and 0.98 dB, respectively, and decreases the LPIPS and DISTS by 0.042 and 0.009, respectively. Clearly, LDDPM can generate high-quality, diverse images that strongly align with HR images.

\begin{figure}[H]
	\centering
	\includegraphics[width=200px]{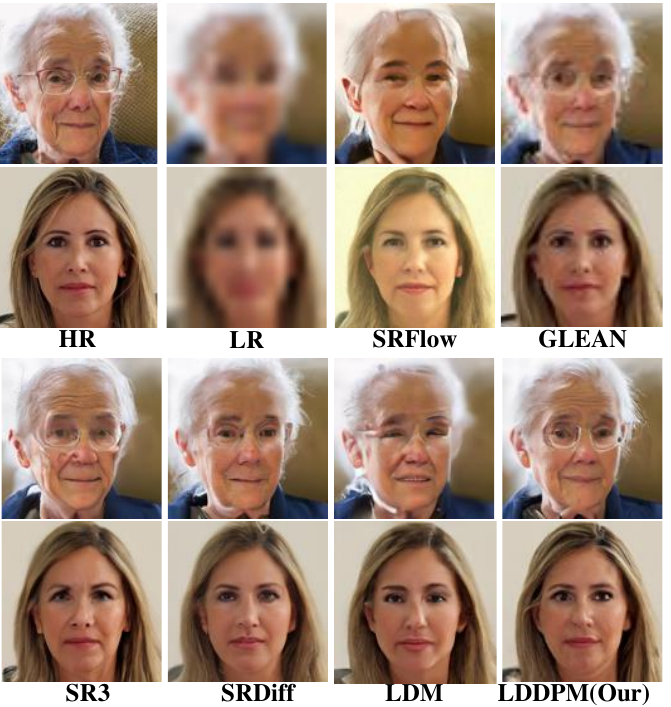}
	\caption{Visualization results of different models on the CelebA dataset (×8, LR image size of 20)}
	\label{fig_6}
\end{figure}

We visualized some model results on the CelebA dataset, as shown in Figure \ref{fig_6}. The elderly and female images reconstructed using the SRFlow model, which is based on normalized flow, are overly smooth. The images reconstructed using the GAN-based GLEAN model were relatively blurry, leading to an unrealistic perception. In contrast, the elderly and female images inferred by the DDPM-based SR3, SRDiff, and LDM models exhibit over-inference, such as the women's expressions and glasses of the elderly. Compared with other models, wrinkles on the forehead of the elderly and the hair of women reconstructed by LDDPM appear more natural, with rich details and textures.

In addition, we compared the model parameters and prediction times. As shown in Figure \ref{fig_6_1}, our LDDPM model has fewer parameters and diffusion steps and requires less time for sampling. Although LDDPM uses more model parameters than the SRDiff model, it requires fewer diffusion steps and less time than SRDiff, indicating that LDDPM can achieve better performance at a smaller computational cost. Furthermore, when training on the CelebA dataset, LDDPM required only approximately 20 h to converge, whereas SRDiff required 34 h, SR3 required 40 h, and LDM required much longer (approximately one week) to converge. These outcomes indicate that LDDPM is efficient in training.

As shown in Figure \ref{fig_7}, we calculated the pixel details of the facial features in the HR images reconstructed by LDDPM, SR3, LDM, and SRDiff using histogram and PSNR metrics. The grayscale histogram of LDDPM is smoother and the PSNR is higher, indicating that LDDPM not only generates images of higher quality but also preserves more facial details.

\begin{table}[H]
	\caption{\label{tab:table3}Quantitative comparison of LDDPM with state-of-the-art models on the CelebA dataset (×8, LR image size of 20)}
	\centering
	\resizebox{1\columnwidth}{!}{
		\setlength{\tabcolsep}{0.5mm}{
			\begin{tabular}{cccccc}
				\toprule
				\textbf{Models} & \multicolumn{1}{c}{PSNR$\uparrow$(dB)} & \multicolumn{1}{c}{SSIM$\uparrow$} & \multicolumn{1}{c}{LPIPS$\downarrow$} & \multicolumn{1}{c}{LR-PSNR$\uparrow$} & \multicolumn{1}{c}{DISTS$\downarrow$} \\
				\midrule
				ESRGAN (ECCV,2018) \cite{DBLP:conf/eccv/WangYWGLDQL18} & 23.24 & 0.6645 & 0.115 &39.91 & 0.111 \\
				GLEAN (CVPR,2022) & 23.90 & 0.6856& 0.135 & 50.45 & 0.109 \\
				ProgFSR (arXiv,2019) \cite{DBLP:conf/bmvc/KimKKK19} & 24.21 & 0.7224 & 0.148 & 48.21 & 0.118 \\
				SRFlow (ECCV,2020) \cite{DBLP:journals/ijon/LiYCCFXLC22} & 25.32 & 0.7245 & 0.110  & 50.58 & 0.114 \\
				SRDiff (NC,2022) & 25.38 & 0.7421 & 0.106 & 52.34 & 0.091 \\
				SR3 (T-PAMI,2022) & 24.92 & 0.7095 & 0.089 & 51.56 & 0.169 \\
				LDM (CVPR,2022) & 25.45 & 0.7232 & 0.104 & 52.23 & 0.083 \\
				LDDPM & \textbf{26.07} & \textbf{0.7652} & \textbf{0.062} & \textbf{53.21} & \textbf{0.074} \\
				\bottomrule
			\end{tabular}%
		}
	}
\end{table}
\begin{figure}[H]
	\centering
	\begin{tikzpicture}
		\begin{axis}[
			ybar,
			symbolic x coords={{LDDPM(93M)}, {SRDiff(52M)}, {SR3(98M)}, {LDM(1066.24M)}},
			xtick=data,
			ylabel={Average Sampling time (s)},
			bar width=0.8cm,
			ymin=0,
			ymax=130,
			ymajorgrids=true,
			grid style=dashed,
			nodes near coords={
				\pgfmathprintnumber{\pgfplotspointmeta} s\\[-0.5ex]
				(T=\ifnum\coordindex=0 1000 \else\ifnum\coordindex=1 2000\else\ifnum\coordindex=2 2000\else\ifnum\coordindex=3 20000\fi\fi\fi\fi)
			},
			nodes near coords style={align=center, anchor=south},
			xticklabel style={align=center, anchor=north, below=3pt, font=\footnotesize},
			every node near coord/.append style={font=\footnotesize, inner sep=0pt},
			axis lines*=left
			,
			]
			\addplot coordinates {(LDDPM(93M), 45) (SRDiff(52M), 93) (SR3(98M), 112) (LDM(1066.24M),116)};
		\end{axis}
	\end{tikzpicture}
	\caption{Comparison of sampling times and diffusion steps for different models on the CelebA dataset (×8, LR image size of 20) }
	\label{fig_6_1}
\end{figure}
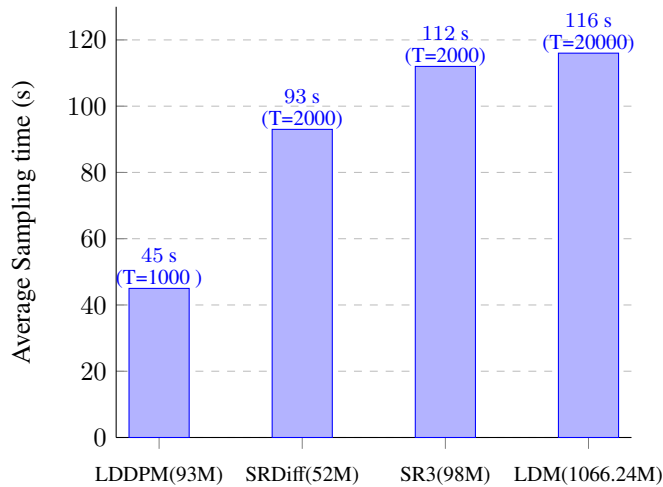
\begin{figure}[H]
	\centering
	\includegraphics[width=\linewidth]{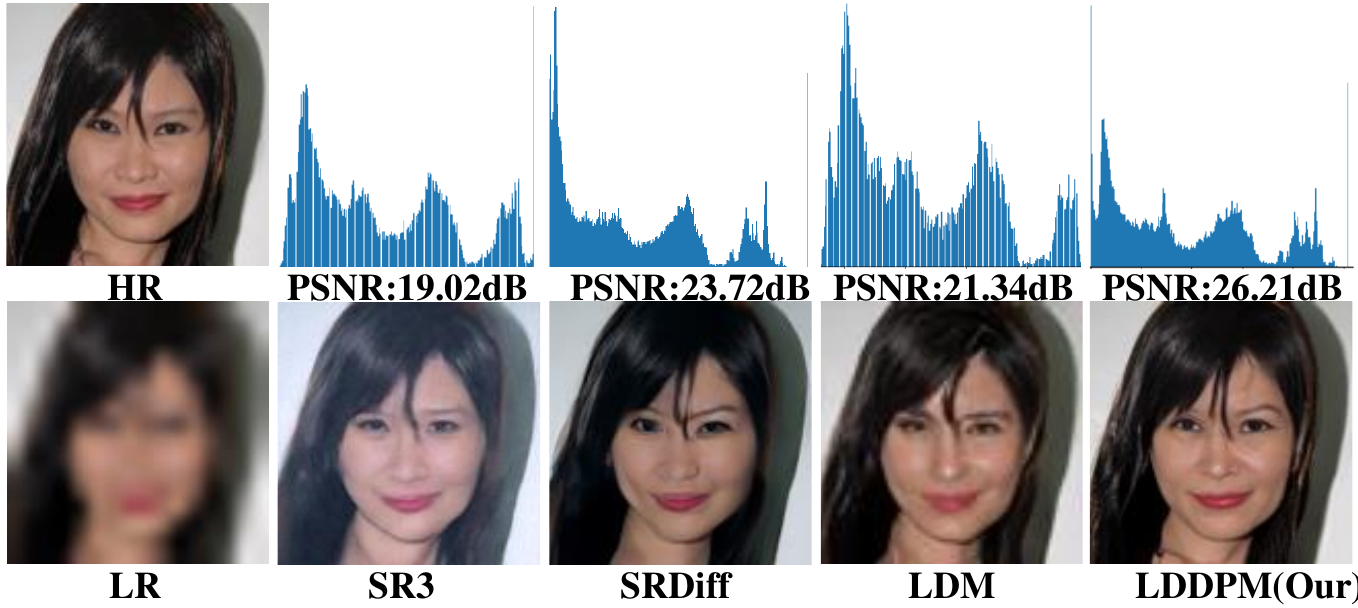}
	\caption{Grayscale histograms and PSNR metrics of the pixel features of the different models displayed on the CelebA dataset}
	\label{fig_7}
\end{figure}
\subsection{Ablation Experiments}
To verify the effectiveness of the added modules in LDDPM, we conducted extensive ablation experiments using our proposed modules on the CelebA dataset.

\textbf{Condition Encode: }In the context of conditional encoding, we define three models and discuss the impact of conditional encoding on LDDPM. The first model (V1) uses conditional encoding by stacking LR and noise images from each stage, thereby reconstructing HR images. The second model (V2) extends V1 by incorporating conditional encoding based on an adaptive multi-head attention mechanism. The third model (V3) builds on V2 by integrating conditional encoding based on the CVAE. As indicated in Table \ref{tab:table4}, the V2 model, which includes the adaptive multi-head attention mechanism, improves the PSNR and SSIM values by 1.05 dB and 0.0123, respectively, over V1 and decreases LPIPS and DISTS by 0.035 and 0.111, respectively. The adaptive multi-head attention mechanism can provide more conditional features to guide the model in learning the probability distribution of HR images, thereby achieving consistency between the reconstructed and actual HR. Furthermore, the V3 model with the CVAE included, when compared with V2, shows an increase in PSNR and SSIM by 1.14 dB and 0.0083, respectively, and a decrease in LPIPS and DISTS by 0.018 and 0.008, respectively. By incorporating CVAE into LDDPM, the model can learn more latent conditional features from the LR images. This not only further narrows the solution space of HR images but also constrains the feature information within the image space.

\begin{figure}[H]
	\centering
	\includegraphics[width=\linewidth]{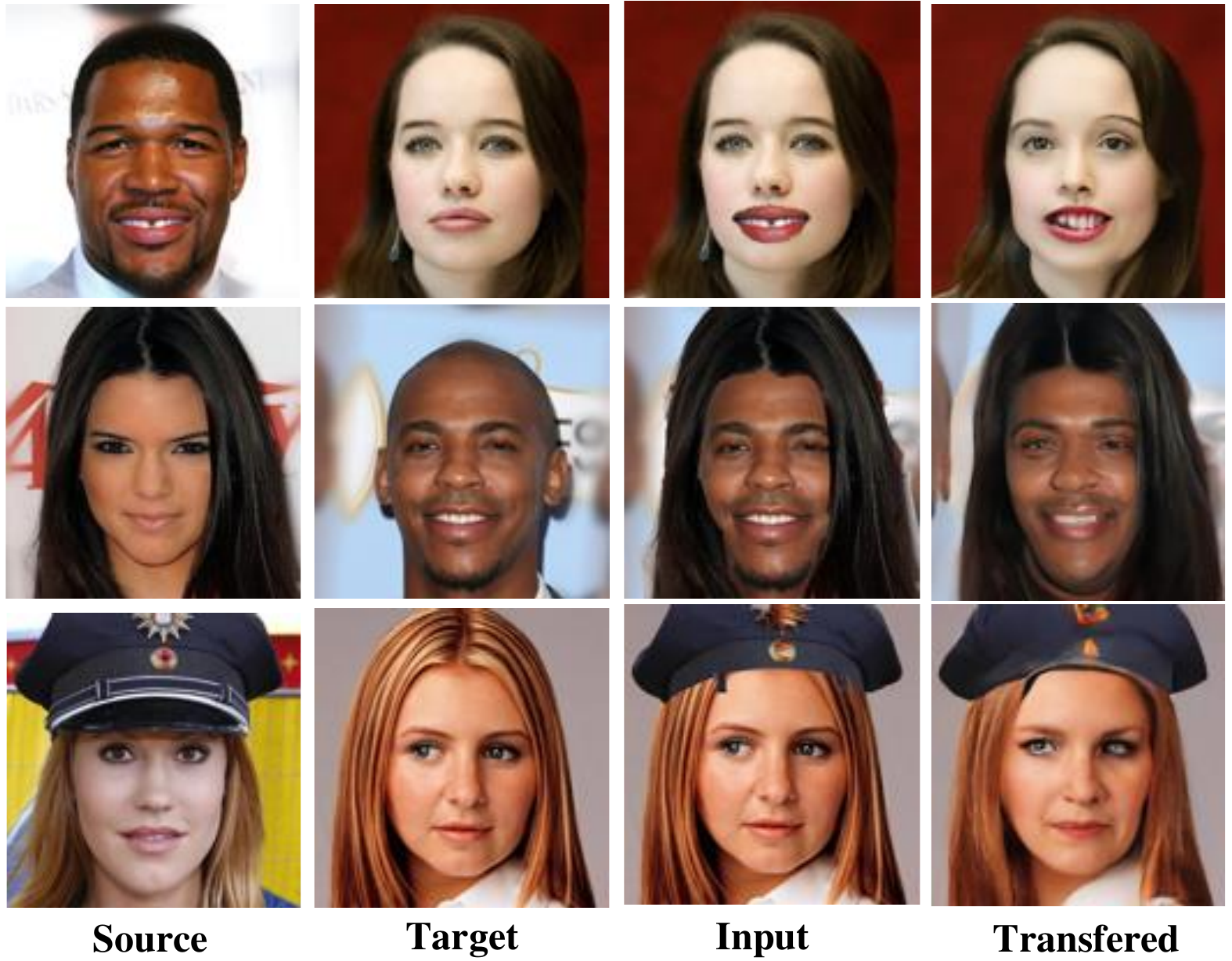}
	\caption{LDDOM integrating the source image with the content in the target image to generate a new image}
	\label{fig_8_1}
\end{figure}

Moreover, within the conditional encoding of LDDPM, we can manipulate the model reconstruction process by simply inserting multiple images. As shown in Figure \ref{fig_8_1}, we fuse the mouth of a black person, a woman's hair, and a woman's hat from the source images with the target image, thereby generating a style transfer image that appears more natural than the original synthetic image.

\begin{figure}[H]
	\centering
	\includegraphics[width=\linewidth]{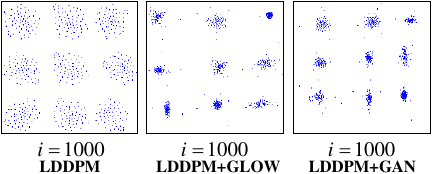}
	\caption{Visualization of feature sampling by LDDPM with the same number of steps}
	\label{fig_8}
\end{figure}

\textbf{Model Optimization based on Glow and GAN: }The second row of Table \ref{tab:table5} indicates that with the inclusion of Glow in LDDPM, the PSNR and SSIM increase by 1.03 dB and 0.0319, respectively, and the LPIPS and DISTS decrease by 0.021 and 0.034, respectively. The inclusion of Glow in LDDPM allows it to capture more complex noise data distributions. The third row of Table \ref{tab:table5} shows that with the inclusion of GAN in LDDPM, the PSNR and SSIM increase by 0.87 dB and 0.0097, respectively, and the LPIPS and DISTS decrease by 0.026 and 0.019, respectively. The multimodal distribution learned by the GAN enables the LDDPM to reconstruct more realistic HR images during the inverse process. In Figure \ref{fig_8}, we visualize the features extracted by LDDPM with the inclusion of Glow and GAN. Compared with the original LDDPM, that with Glow and GAN can learn better data distributions with relatively fewer sampling steps.

\begin{figure}[H]
	\centering
	\includegraphics[width=\linewidth]{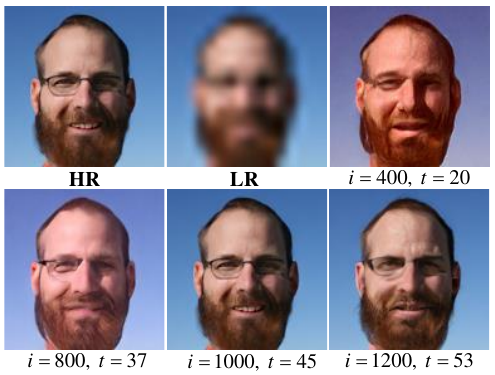}
	\caption{LDDPM generates HR images for different steps, where $t$ is the time to generate HR images in seconds.}
	\label{fig_9}
\end{figure}

\textbf{ Optimization of Experimental Hyperparameters: }To explore the effects of the total diffusion steps and loss functions on LDDPM, we also performed ablation experiments on the hyperparameters. As shown in Figure \ref{fig_9}, the image quality improved with an increase in the total number of diffusion steps. However, a larger number of total diffusion steps would slow down the training and inference of the model; therefore, we opted for a default parameter setting. Finally, we compared the effects of CL and SL on the experimental results. From Table \ref{tab:table6}, it can be seen that adding CL to LDDPM may result in a 0.002 higher DISTS compared with the first row of Table \ref{tab:table6}, but PSNR and SSIM increase by 0.56 dB and 0.0049, respectively, while LPIPS decreases by 0.003. When SL was added to LDDPM, compared with the second row of Table \ref{tab:table6}, PSNR and SSIM increased by 0.18 dB and 0.0064, respectively, whereas LPIPS and DISTS decreased by 0.003 and 0.013, respectively. The above results demonstrate that adding CL and SL to LDDPM can better guide it in learning more image feature information, thereby enabling more stable training.

\begin{table}[!ht]
	\caption{\label{tab:table4}Comparison of metrics with the inclusion of the conditional encoding module in LDDPM, with the best results displayed in bold}
	\centering
	\begin{tabular}{rrrrr}
		\toprule
		\textbf{Models} & \multicolumn{1}{c}{PSNR$\uparrow$(dB)} & \multicolumn{1}{c}{SSIM$\uparrow$} & \multicolumn{1}{c}{LPIPS$\downarrow$} & \multicolumn{1}{c}{DISTS$\downarrow$} \\
		\midrule
		V1    & 23.10  & 0.7109 & 0.127 & 0.218 \\
		V2    & 24.15 & 0.7232 & 0.092 & 0.107 \\
		V3    & \textbf{25.29} & \textbf{0.7315} & \textbf{0.074} & \textbf{0.099} \\
		\bottomrule
	\end{tabular}%
\end{table}
\begin{table}[!ht]
	\caption{\label{tab:table5}Comparisons of metrics with the inclusion of Glow and GAN in LDDPM, with the best results highlighted in bold}
	\centering
	\setlength{\tabcolsep}{0.8mm}{
		\begin{tabular}{ccccc}
			\toprule
			\textbf{Models} & \multicolumn{1}{c}{PSNR$\uparrow$(dB)} & \multicolumn{1}{c}{SSIM$\uparrow$} & \multicolumn{1}{c}{LPIPS$\downarrow$} & \multicolumn{1}{c}{DISTS$\downarrow$} \\
			\midrule
			LDDPM & 23.43 & 0.7123 & 0.115 & 0.132 \\
			LDDPM+Glow & 24.46 & 0.74.42 & 0.094 & 0.098 \\
			LDDPM+Glow+GAN & \textbf{25.33} & \textbf{0.7539} & \textbf{0.068} & \textbf{0.079} \\
			\bottomrule
		\end{tabular}%
	}
\end{table}
\begin{table}[!ht]
	\caption{\label{tab:table6}Comparisons of metrics with the inclusion of CL and SL in LDDPM, with the best results highlighted in bold}
	\centering
	\setlength{\tabcolsep}{1mm}{
		\begin{tabular}{ccccc}
			\toprule
			\textbf{Models} & \multicolumn{1}{c}{PSNR$\uparrow$(dB)} & \multicolumn{1}{c}{SSIM$\uparrow$} & \multicolumn{1}{c}{LPIPS$\downarrow$} & \multicolumn{1}{c}{DISTS$\downarrow$} \\
			\midrule
			LDDPM & 25.33 & 0.7539 & 0.068 & 0.079 \\
			LDDPM+CL & 25.89 & 0.7588 & 0.065 & 0.081 \\
			LDDPM+CL+SL & \textbf{26.07} & \textbf{0.7652} & \textbf{0.062} & \textbf{0.074} \\
			\bottomrule
		\end{tabular}%
	}
	
\end{table}
\subsection{Experimental Comparison of Real Datasets}
\begin{figure}[H]
	\centering
	\includegraphics[width=\linewidth]{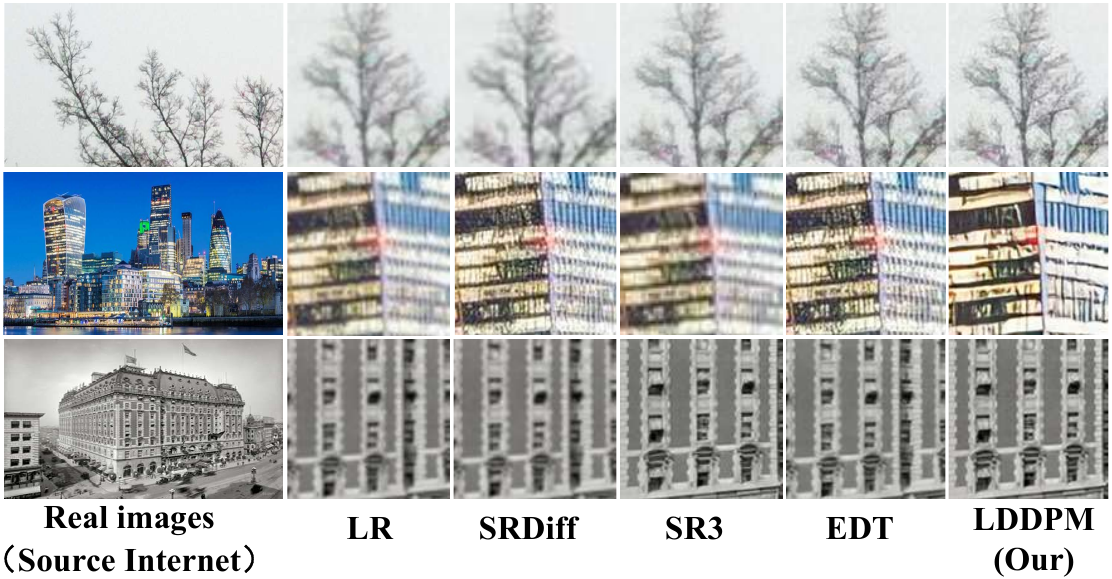}
	\caption{LDDPM generates HR images under different steps, where $t$ is the time to generate HR images in seconds.}
	\label{fig_10}
\end{figure}

To evaluate the performance of the LDDPM more comprehensively, we collected LR images from the real world. As shown in Figure \ref{fig_10}, the quality of the reconstructed HR images from LDDPM was better than that of the reconstructed SR3, EDF, and SRDiff. Specifically, the images reconstructed using SR3, EDF, and SRDiff in Figure \ref{fig_10} are blurred and have missing details and textures. In contrast, the LDDPM can reconstruct not only clear images but also images with complete details and textures. Experiments on real-world datasets demonstrated that LDDPM generalizes well and can be applied to SISR tasks in natural environments. 

\section{Conclusion}
We design LDDPM to address the "one-to-many" uncertainty of the SISR task. This solves the problems of missing details and textures in reconstructed HR images, slow sampling speed of the model, and ineffective use of degraded images in existing image super-resolution reconstruction methods. LDDPM mainly uses Markov chains to convert HR images into simple Gaussian probability distributions and then uses the inverse diffusion process to reconstruct HR images gradually. We used a conditional encoder in the forward and reverse processes of LDDPM. The conditional encoder encodes the LR image using an adaptive multi-headed attention mechanism and CVAE, which significantly constrains the solution space of the reconstructed image. In addition, to accelerate the convergence speed and ensure stable training of LDDPM, we added a normalized flow and multimodal adversarial training to the model. These methods utilize a complex distribution to model each denoising process, enabling the model to learn the probability distribution of more complex HR images efficiently and significantly reduce the number of diffusion steps of LDDPM. Extensive experiments have demonstrated that LDDPM can better utilize LR image feature information to generate HR images with better perceptual quality in a smaller number of diffusion steps.

Although our work is meaningful, there are areas that require improvement. The first is the high sampling count in LDDPM and its possible reduction through score-based methods. Moreover, adding various types of noise (Gaussian white, JPEG compressed, and reversed ISP-generated sensor noise) during LDDPM's forward diffusion can potentially enhance model robustness. Moving forward, our research will focus on the following aspects:
\begin{enumerate}{}{}
	\item{ Exploring score-based DDPM methods to significantly decrease LDDPM's sampling steps.}
	\item{ Improving LDDPM's HR image reconstruction by introducing diverse noise types during forward diffusion.}
\end{enumerate}

\subsection*{Acknowledgements}
This work is supported by General Project of Guangxi Science and Technology Major Project (AA19254016), Beihai City Science and Technology Planning Project (202082033), Beihai City Science and Technology Planning Project (202082023), Guangxi Graduate Student Innovation Project (YCSW2021174). 

\subsection*{Declaration of competing interest}

The authors declare that they have no known competing financial interests or personal relationships that could have appeared to
influence the work reported in this paper.

\bibliographystyle{CVMbib}
\bibliography{Manuscript}

\begin{thebibliography}{10}
\expandafter\ifx\csname urlstyle\endcsname\relax
  \providecommand{\doi}[1]{doi:\discretionary{}{}{}#1}\else
  \providecommand{\doi}{doi:\discretionary{}{}{}\begingroup
  \urlstyle{rm}\Url}\fi

\bibitem{DBLP:journals/tip/ChengFLZSW22}
Cheng L, Fang P, Liang Y, Zhang L, Shen C, Wang H. {TSGB:} Target-Selective
  Gradient Backprop for Probing {CNN} Visual Saliency. \emph{{IEEE} Trans.
  Image Process.}, 2022, 31: 2529--2540.

\bibitem{jiang2023sphere}
Jiang D, Jin Y, Zhang FL, Zhu Z, Zhang Y, Tong R, Tang M. Sphere Face Model: A
  3D morphable model with hypersphere manifold latent space using joint 2D/3D
  training. \emph{Computational Visual Media}, 2023, 9(2): 279--296.

\bibitem{DBLP:journals/tii/WangXLXX22}
Wang M, Xu Z, Liu X, Xiong J, Xie W. Perceptually Quasi-Lossless Compression of
  Screen Content Data Via Visibility Modeling and Deep Forecasting.
  \emph{{IEEE} Trans. Ind. Informatics}, 2022, 18(10): 6865--6875.

\bibitem{chen2023towards}
Chen S, Wang J, Pan W, Gao S, Wang M, Lu X. Towards uniform point distribution
  in feature-preserving point cloud filtering. \emph{Computational Visual
  Media}, 2023, 9(2): 249--263.

\bibitem{DBLP:journals/vc/MaT21}
Ma T, Tian W. Back-projection-based progressive growing generative adversarial
  network for single image super-resolution. \emph{Vis. Comput.}, 2021, 37(5):
  925--938.

\bibitem{DBLP:journals/spic/KarimiT21}
Karimi N, Taban MR. A convex variational method for super resolution of {SAR}
  image with speckle noise. \emph{Signal Process. Image Commun.}, 2021, 90:
  116061.

\bibitem{DBLP:conf/cvpr/ZhouHGZ21}
Zhou H, Huang C, Gao S, Zhuang X. VSpSR: Explorable Super-Resolution via
  Variational Sparse Representation. In \emph{{IEEE} Conference on Computer
  Vision and Pattern Recognition Workshops, {CVPR} Workshops 2021, virtual,
  June 19-25, 2021}, 2021, 373--381.

\bibitem{DBLP:journals/tgrs/ShiHHCHD22}
Shi Y, Han L, Han L, Chang S, Hu T, Dancey D. A Latent Encoder Coupled
  Generative Adversarial Network {(LE-GAN)} for Efficient Hyperspectral Image
  Super-Resolution. \emph{{IEEE} Trans. Geosci. Remote. Sens.}, 2022, 60:
  1--19.

\bibitem{DBLP:conf/iccv/LiangL0DGT21}
Liang J, Lugmayr A, Zhang K, Danelljan M, Gool LV, Timofte R. Hierarchical
  Conditional Flow: {A} Unified Framework for Image Super-Resolution and Image
  Rescaling. In \emph{2021 {IEEE/CVF} International Conference on Computer
  Vision, {ICCV} 2021, Montreal, QC, Canada, October 10-17, 2021}, 2021,
  4056--4065.

\bibitem{DBLP:conf/cvpr/LiuSW21}
Liu Z, Siu W, Wang L. Variational AutoEncoder for Reference Based Image
  Super-Resolution. In \emph{{IEEE} Conference on Computer Vision and Pattern
  Recognition Workshops, {CVPR} Workshops 2021, virtual, June 19-25, 2021},
  2021, 516--525.

\bibitem{DBLP:journals/ijon/LiYCCFXLC22}
Li H, Yang Y, Chang M, Chen S, Feng H, Xu Z, Li Q, Chen Y. SRDiff: Single image
  super-resolution with diffusion probabilistic models. \emph{Neurocomputing},
  2022, 479: 47--59.

\bibitem{DBLP:journals/corr/abs-2111-14822}
Gu S, Chen D, Bao J, Wen F, Zhang B, Chen D, Yuan L, Guo B. Vector Quantized
  Diffusion Model for Text-to-Image Synthesis. In \emph{{IEEE/CVF} Conference
  on Computer Vision and Pattern Recognition, {CVPR} 2022, New Orleans, LA,
  USA, June 18-24, 2022}, 2022, 10686--10696.

\bibitem{DBLP:conf/icml/KimKS21}
Kim J, Kong J, Son J. Conditional Variational Autoencoder with Adversarial
  Learning for End-to-End Text-to-Speech. In M~Meila, T~Zhang, editors,
  \emph{Proceedings of the 38th International Conference on Machine Learning,
  {ICML} 2021, 18-24 July 2021, Virtual Event}, volume 139 of \emph{Proceedings
  of Machine Learning Research}, 2021, 5530--5540.

\bibitem{DBLP:journals/tcsv/ZhangLWPMYY22}
Zhang J, Long C, Wang Y, Piao H, Mei H, Yang X, Yin B. A Two-Stage Attentive
  Network for Single Image Super-Resolution. \emph{{IEEE} Trans. Circuits Syst.
  Video Technol.}, 2022, 32(3): 1020--1033.

\bibitem{DBLP:conf/cvpr/DaiCZXZ19}
Dai T, Cai J, Zhang Y, Xia S, Zhang L. Second-Order Attention Network for
  Single Image Super-Resolution. In \emph{{IEEE} Conference on Computer Vision
  and Pattern Recognition, {CVPR} 2019, Long Beach, CA, USA, June 16-20, 2019},
  2019, 11065--11074.

\bibitem{DBLP:conf/eccv/NiuWRZYWZCS20}
Niu B, Wen W, Ren W, Zhang X, Yang L, Wang S, Zhang K, Cao X, Shen H. Single
  Image Super-Resolution via a Holistic Attention Network. In \emph{Computer
  Vision - {ECCV} 2020 - 16th European Conference, Glasgow, UK, August 23-28,
  2020, Proceedings, Part {XII}}, volume 12357 of \emph{Lecture Notes in
  Computer Science}, 2020, 191--207.

\bibitem{DBLP:conf/nips/ZhouZZL20}
Zhou S, Zhang J, Zuo W, Loy CC. Cross-Scale Internal Graph Neural Network for
  Image Super-Resolution. In \emph{Advances in Neural Information Processing
  Systems 33: Annual Conference on Neural Information Processing Systems 2020,
  NeurIPS 2020, December 6-12, 2020, virtual}, volume~33, 2020, 3499--3509.

\bibitem{DBLP:conf/eccv/WangYWGLDQL18}
Wang X, Yu K, Wu S, Gu J, Liu Y, Dong C, Qiao Y, Loy CC. {ESRGAN:} Enhanced
  Super-Resolution Generative Adversarial Networks. In L~Leal{-}Taix{\'{e}},
  S~Roth, editors, \emph{Computer Vision - {ECCV} 2018 Workshops - Munich,
  Germany, September 8-14, 2018, Proceedings, Part {V}}, volume 11133 of
  \emph{Lecture Notes in Computer Science}, 2018, 63--79.

\bibitem{DBLP:conf/cvpr/ChanWXGL21}
Chan KCK, Wang X, Xu X, Gu J, Loy CC. {GLEAN:} Generative Latent Bank for
  Large-Factor Image Super-Resolution. In \emph{{IEEE} Conference on Computer
  Vision and Pattern Recognition, {CVPR} 2021, virtual, June 19-25, 2021},
  2021, 14245--14254.

\bibitem{9815113}
Liu Z, Li Z, Wu X, Liu Z, Chen W. {DSRGAN:} Detail Prior-Assisted Perceptual
  Single Image Super-Resolution via Generative Adversarial Networks.
  \emph{{IEEE} Trans. Circuits Syst. Video Technol.}, 2022, 32(11): 7418--7431.

\bibitem{DBLP:journals/corr/abs-2006-05218}
Gatopoulos I, Stol M, Tomczak JM. Super-resolution Variational Auto-Encoders.
  \emph{CoRR}, 2020, abs/2006.05218.

\bibitem{liu2020photo}
Liu Z, Siu W, Chan Y. Photo-Realistic Image Super-Resolution via Variational
  Autoencoders. \emph{{IEEE} Trans. Circuits Syst. Video Technol.}, 2021,
  31(4): 1351--1365.

\bibitem{xiang2021learning}
Xiang X, Zhu L, Li J, Wang Y, Huang T, Tian Y. Learning Super-Resolution
  Reconstruction for High Temporal Resolution Spike Stream. \emph{{IEEE} Trans.
  Circuits Syst. Video Technol.}, 2023, 33(1): 16--29.

\bibitem{DBLP:conf/cvpr/JoYK21}
Jo Y, Yang S, Kim SJ. SRFlow-DA: Super-Resolution Using Normalizing Flow With
  Deep Convolutional Block. In \emph{{IEEE} Conference on Computer Vision and
  Pattern Recognition Workshops, {CVPR} Workshops 2021, virtual, June 19-25,
  2021}, 2021, 364--372.

\bibitem{DBLP:journals/corr/abs-2104-07636}
Saharia C, Ho J, Chan W, Salimans T, Fleet DJ, Norouzi M. Image
  Super-Resolution via Iterative Refinement. \emph{{IEEE} Trans. Pattern Anal.
  Mach. Intell.}, 2023, 45(4): 4713--4726.

\bibitem{DBLP:journals/corr/abs-2208-01864}
Ryu D, Ye JC. Pyramidal Denoising Diffusion Probabilistic Models. \emph{CoRR},
  2022, abs/2208.01864.

\bibitem{DBLP:journals/corr/abs-2303-09472}
Xia B, Zhang Y, Wang S, Wang Y, Wu X, Tian Y, Yang W, Gool LV. DiffIR:
  Efficient Diffusion Model for Image Restoration. \emph{CoRR}, 2023,
  abs/2303.09472.

\bibitem{DBLP:journals/corr/abs-2212-00490}
Wang Y, Yu J, Zhang J. Zero-Shot Image Restoration Using Denoising Diffusion
  Null-Space Model. \emph{CoRR}, 2022, abs/2212.00490.

\bibitem{RombachBLEO22}
Rombach R, Blattmann A, Lorenz D, Esser P, Ommer B. High-Resolution Image
  Synthesis with Latent Diffusion Models. In \emph{{IEEE/CVF} Conference on
  Computer Vision and Pattern Recognition, {CVPR} 2022, New Orleans, LA, USA,
  June 18-24, 2022}, 2022, 10674--10685.

\bibitem{DBLP:journals/corr/abs-2006-11239}
Ho J, Jain A, Abbeel P. Denoising Diffusion Probabilistic Models. In
  \emph{Advances in Neural Information Processing Systems 33: Annual Conference
  on Neural Information Processing Systems 2020, NeurIPS 2020, December 6-12,
  2020, virtual}, volume~33, 2020, 6840--6851.

\bibitem{DBLP:conf/icml/NicholD21}
Nichol AQ, Dhariwal P. Improved denoising diffusion probabilistic models. In
  \emph{International Conference on Machine Learning}, 2021, 8162--8171.

\bibitem{DBLP:journals/access/QinYWWLW21}
Qin Q, Yan J, Wang X, Wang Q, Li M, Wang Y. ETDNet: An Efficient Transformer
  Deraining Model. \emph{{IEEE} Access}, 2021, 9: 119881--119893.

\bibitem{DBLP:conf/www/LiangKHJ18}
Liang D, Krishnan RG, Hoffman MD, Jebara T. Variational Autoencoders for
  Collaborative Filtering. In P~Champin, F~Gandon, M~Lalmas, PG~Ipeirotis,
  editors, \emph{Proceedings of the 2018 World Wide Web Conference on World
  Wide Web, {WWW} 2018, Lyon, France, April 23-27, 2018}, 2018, 689--698.

\bibitem{DBLP:conf/nips/KingmaD18}
Kingma DP, Dhariwal P. Glow: Generative Flow with Invertible 1x1 Convolutions.
  In S~Bengio, HM~Wallach, H~Larochelle, K~Grauman, N~Cesa{-}Bianchi,
  R~Garnett, editors, \emph{Advances in Neural Information Processing Systems
  31: Annual Conference on Neural Information Processing Systems 2018, NeurIPS
  2018, December 3-8, 2018, Montr{\'{e}}al, Canada}, volume~31, 2018,
  10236--10245.

\bibitem{DBLP:conf/iclr/XiaoKV22}
Xiao Z, Kreis K, Vahdat A. Tackling the Generative Learning Trilemma with
  Denoising Diffusion GANs. In \emph{The Tenth International Conference on
  Learning Representations, {ICLR} 2022, Virtual Event, April 25-29, 2022},
  2022, https://openreview.net/forum?id=JprM0p--q0Co.

\bibitem{DBLP:conf/cvpr/LedigTHCCAATTWS17}
Ledig C, Theis L, Huszar F, Caballero J, Cunningham A, Acosta A, Aitken AP,
  Tejani A, Totz J, Wang Z, Shi W. Photo-Realistic Single Image
  Super-Resolution Using a Generative Adversarial Network. In \emph{2017 {IEEE}
  Conference on Computer Vision and Pattern Recognition, {CVPR} 2017, Honolulu,
  HI, USA, July 21-26, 2017}, 2017, 105--114.

\bibitem{DBLP:conf/cvpr/ParkL19}
Park DY, Lee KH. Arbitrary Style Transfer With Style-Attentional Networks. In
  \emph{{IEEE} Conference on Computer Vision and Pattern Recognition, {CVPR}
  2019, Long Beach, CA, USA, June 16-20, 2019}, 2019, 5880--5888.

\bibitem{DBLP:journals/pami/KarrasLA21}
Karras T, Laine S, Aila T. A Style-Based Generator Architecture for Generative
  Adversarial Networks. \emph{{IEEE} Trans. Pattern Anal. Mach. Intell.}, 2021,
  43(12): 4217--4228.

\bibitem{liu2018large}
Liu Z, Luo P, Wang X, Tang X. Large-scale celebfaces attributes (celeba)
  dataset. \emph{Retrieved August}, 2018, 15(2018): 11.

\bibitem{DBLP:conf/cvpr/AgustssonT17}
Agustsson E, Timofte R. {NTIRE} 2017 Challenge on Single Image
  Super-Resolution: Dataset and Study. In \emph{2017 {IEEE} Conference on
  Computer Vision and Pattern Recognition Workshops, {CVPR} Workshops 2017,
  Honolulu, HI, USA, July 21-26, 2017}, 2017, 1122--1131.

\bibitem{DBLP:conf/cvpr/LimSKNL17}
Lim B, Son S, Kim H, Nah S, Lee KM. Enhanced Deep Residual Networks for Single
  Image Super-Resolution. In \emph{2017 {IEEE} Conference on Computer Vision
  and Pattern Recognition Workshops, {CVPR} Workshops 2017, Honolulu, HI, USA,
  July 21-26, 2017}, 2017, 1132--1140.

\bibitem{DBLP:conf/iccv/0008LGT21}
Zhang K, Liang J, Gool LV, Timofte R. Designing a Practical Degradation Model
  for Deep Blind Image Super-Resolution. In \emph{2021 {IEEE/CVF} International
  Conference on Computer Vision, {ICCV} 2021, Montreal, QC, Canada, October
  10-17, 2021}, 2021, 4771--4780.

\bibitem{DBLP:conf/iclr/LoshchilovH19}
Loshchilov I, Hutter F. Decoupled Weight Decay Regularization. In \emph{7th
  International Conference on Learning Representations, {ICLR} 2019, New
  Orleans, LA, USA, May 6-9, 2019}, 2019,
  https://openreview.net/forum?id=Bkg6RiCqY7.

\bibitem{DBLP:conf/eccv/ZhangLLWZF18}
Zhang Y, Li K, Li K, Wang L, Zhong B, Fu Y. Image Super-Resolution Using Very
  Deep Residual Channel Attention Networks. In V~Ferrari, M~Hebert,
  C~Sminchisescu, Y~Weiss, editors, \emph{Computer Vision - {ECCV} 2018 - 15th
  European Conference, Munich, Germany, September 8-14, 2018, Proceedings, Part
  {VII}}, volume 11211 of \emph{Lecture Notes in Computer Science}, 2018,
  294--310.

\bibitem{DBLP:conf/cvpr/MeiFZ21}
Mei Y, Fan Y, Zhou Y. Image Super-Resolution With Non-Local Sparse Attention.
  In \emph{{IEEE} Conference on Computer Vision and Pattern Recognition, {CVPR}
  2021, virtual, June 19-25, 2021}, 2021, 3517--3526.

\bibitem{DBLP:conf/iccvw/LiangCSZGT21}
Liang J, Cao J, Sun G, Zhang K, Gool LV, Timofte R. SwinIR: Image Restoration
  Using Swin Transformer. In \emph{{IEEE/CVF} International Conference on
  Computer Vision Workshops, {ICCVW} 2021, Montreal, BC, Canada, October 11-17,
  2021}, 2021, 1833--1844.

\bibitem{DBLP:journals/corr/abs-2208-11247}
Zhang D, Huang F, Liu S, Wang X, Jin Z. SwinFIR: Revisiting the SwinIR with
  Fast Fourier Convolution and Improved Training for Image Super-Resolution.
  \emph{CoRR}, 2022, abs/2208.11247.

\bibitem{DBLP:journals/corr/abs-2112-10175}
Li W, Lu X, Lu J, Zhang X, Jia J. On Efficient Transformer and Image
  Pre-training for Low-level Vision. \emph{CoRR}, 2021, abs/2112.10175.

\bibitem{DBLP:conf/cvpr/WangYDL18}
Wang X, Yu K, Dong C, Loy CC. Recovering Realistic Texture in Image
  Super-Resolution by Deep Spatial Feature Transform. In \emph{2018 {IEEE}
  Conference on Computer Vision and Pattern Recognition, {CVPR} 2018, Salt Lake
  City, UT, USA, June 18-22, 2018}, 2018, 606--615.

\bibitem{DBLP:conf/cvpr/ZhangGT20}
Zhang K, Gool LV, Timofte R. Deep Unfolding Network for Image Super-Resolution.
  In \emph{2020 {IEEE/CVF} Conference on Computer Vision and Pattern
  Recognition, {CVPR} 2020, Seattle, WA, USA, June 13-19, 2020}, 2020,
  3214--3223.

\bibitem{DBLP:conf/cvpr/MaRCCL020}
Ma C, Rao Y, Cheng Y, Chen C, Lu J, Zhou J. Structure-Preserving Super
  Resolution With Gradient Guidance. In \emph{2020 {IEEE/CVF} Conference on
  Computer Vision and Pattern Recognition, {CVPR} 2020, Seattle, WA, USA, June
  13-19, 2020}, 2020, 7766--7775.

\bibitem{DBLP:conf/aaai/LiZQLL22}
Li W, Zhou K, Qi L, Lu L, Lu J. Best-Buddy GANs for Highly Detailed Image
  Super-resolution. In \emph{Thirty-Sixth {AAAI} Conference on Artificial
  Intelligence, {AAAI} 2022, Thirty-Fourth Conference on Innovative
  Applications of Artificial Intelligence, {IAAI} 2022, The Twelveth Symposium
  on Educational Advances in Artificial Intelligence, {EAAI} 2022 Virtual
  Event, February 22 - March 1, 2022}, volume~34, 2022, 1412--1420.

\bibitem{DBLP:conf/cvpr/ParmarLLT21}
Parmar G, Li D, Lee K, Tu Z. Dual Contradistinctive Generative Autoencoder. In
  \emph{{IEEE} Conference on Computer Vision and Pattern Recognition, {CVPR}
  2021, virtual, June 19-25, 2021}, 2021, 823--832.

\bibitem{DBLP:conf/nips/SinhaSME21}
Sinha A, Song J, Meng C, Ermon S. D2C: Diffusion-Decoding Models for Few-Shot
  Conditional Generation. In \emph{Advances in Neural Information Processing
  Systems}, volume~34, 2021, 12533--12548.

\bibitem{DBLP:conf/nips/VahdatK20}
Vahdat A, Kautz J. NVAE: A Deep Hierarchical Variational Autoencoder. In
  H~Larochelle, M~Ranzato, R~Hadsell, M~Balcan, H~Lin, editors, \emph{Advances
  in Neural Information Processing Systems}, volume~33, 2020, 19667--19679.

\bibitem{DBLP:conf/eccv/LugmayrDGT20}
Lugmayr A, Danelljan M, Gool LV, Timofte R. SRFlow: Learning the
  Super-Resolution Space with Normalizing Flow. In A~Vedaldi, H~Bischof,
  T~Brox, J~Frahm, editors, \emph{Computer Vision - {ECCV} 2020 - 16th European
  Conference, Glasgow, UK, August 23-28, 2020, Proceedings, Part {V}}, volume
  12350 of \emph{Lecture Notes in Computer Science}, 2020, 715--732.

\bibitem{DBLP:conf/aaai/Cao0WGS20}
Cao B, Zhang H, Wang N, Gao X, Shen D. Auto-GAN: Self-Supervised Collaborative
  Learning for Medical Image Synthesis. In \emph{The Thirty-Fourth {AAAI}
  Conference on Artificial Intelligence, {AAAI} 2020, The Thirty-Second
  Innovative Applications of Artificial Intelligence Conference, {IAAI} 2020,
  The Tenth {AAAI} Symposium on Educational Advances in Artificial
  Intelligence, {EAAI} 2020, New York, NY, USA, February 7-12, 2020},
  volume~34, 2020, 10486--10493.

\bibitem{DBLP:conf/iclr/BrockDS19}
Brock A, Donahue J, Simonyan K. Large Scale {GAN} Training for High Fidelity
  Natural Image Synthesis. In \emph{7th International Conference on Learning
  Representations, {ICLR} 2019, New Orleans, LA, USA, May 6-9, 2019}, 2019,
  https://openreview.net/forum?id=B1xsqj09Fm.

\bibitem{DBLP:conf/iclr/MiyatoKKY18}
Miyato T, Kataoka T, Koyama M, Yoshida Y. Spectral Normalization for Generative
  Adversarial Networks. In \emph{6th International Conference on Learning
  Representations, {ICLR} 2018, Vancouver, BC, Canada, April 30 - May 3, 2018,
  Conference Track Proceedings}, 2018,
  https://openreview.net/forum?id=B1QRgziT--.

\bibitem{DBLP:conf/nips/SongE19}
Song Y, Ermon S. Generative Modeling by Estimating Gradients of the Data
  Distribution. In HM~Wallach, H~Larochelle, A~Beygelzimer,
  F~d'Alch{\'{e}}{-}Buc, EB~Fox, R~Garnett, editors, \emph{Advances in Neural
  Information Processing Systems 32: Annual Conference on Neural Information
  Processing Systems 2019, NeurIPS 2019, December 8-14, 2019, Vancouver, BC,
  Canada}, volume~32, 2019, 11895--11907.

\bibitem{DBLP:conf/nips/0011E20}
Song Y, Ermon S. Improved Techniques for Training Score-Based Generative
  Models. In \emph{Advances in Neural Information Processing Systems 33: Annual
  Conference on Neural Information Processing Systems 2020, NeurIPS 2020,
  December 6-12, 2020, virtual}, volume~33, 2020, 12438--12448.

\bibitem{DBLP:conf/bmvc/KimKKK19}
Kim D, Kim M, Kwon G, Kim D. Progressive Face Super-Resolution via Attention to
  Facial Landmark. In \emph{30th British Machine Vision Conference 2019, {BMVC}
  2019, Cardiff, UK, September 9-12, 2019}, 2019, 192.

\end{thebibliography}


\subsection*{Author biography}                                     
\begin{biography}[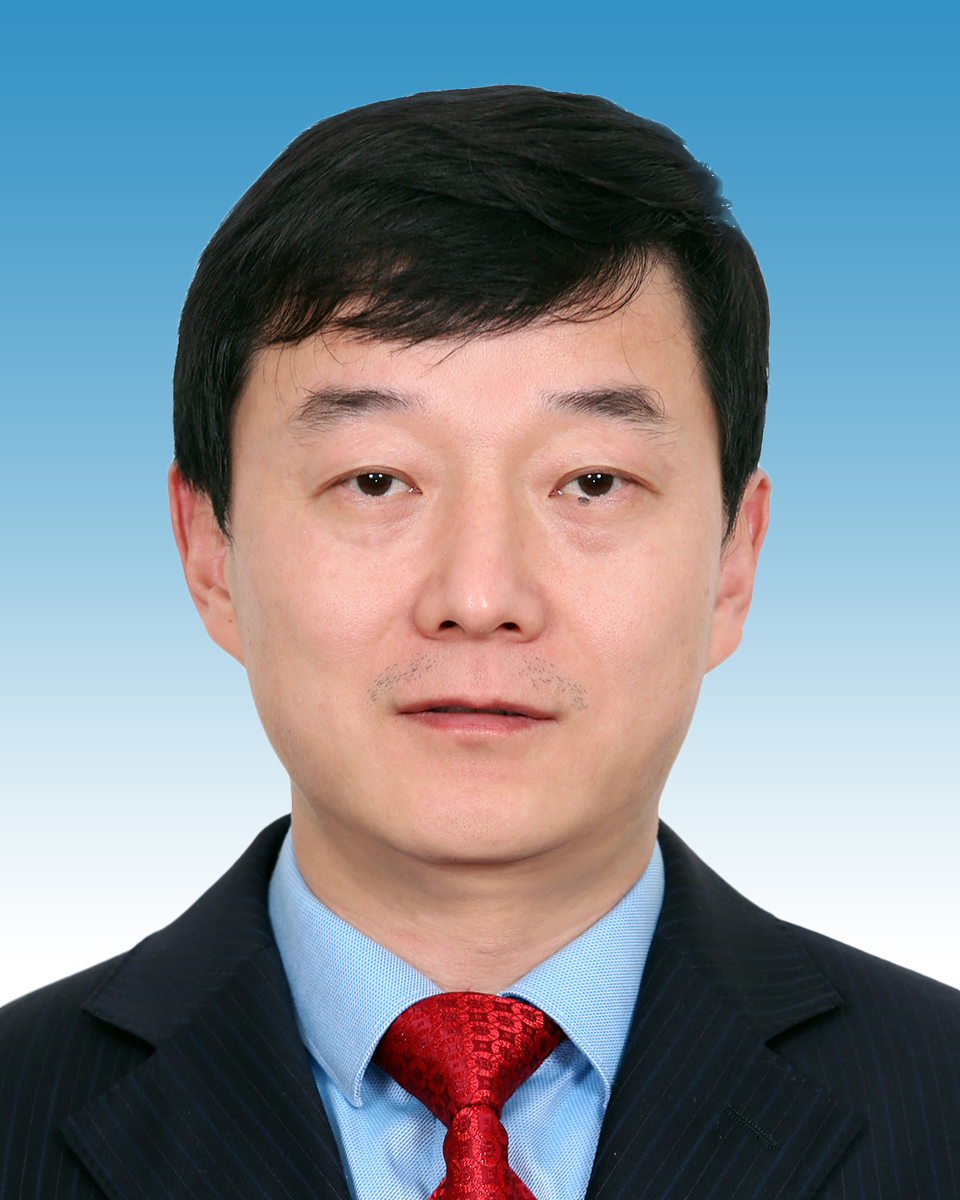]{Xin Wang} (e-mail: 303479506@qq.com) is a professor, and master's supervisor. His main research directions are image processing, network information security, Internet of Things, data mining. He is the main participant of National Natural Science Foundation of China, and the principal investigator of General Project of Guangxi Natural Science Foundation, Guangxi Science and Technology Major Project, Guangxi government planning project and etc.
\end{biography}

\vspace*{2.6em}
\begin{biography}[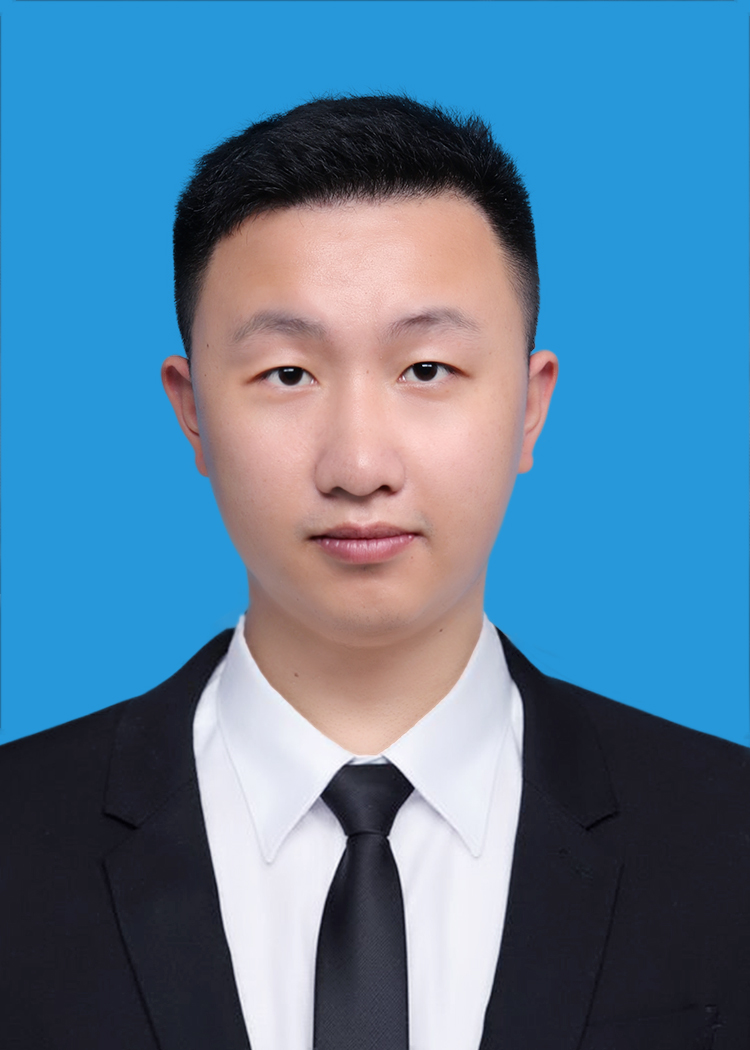]{Jing-Ke Yan} (e-mail: 592499985@qq.com) is currently a postgraduate student in the school of Marine Engineering of Guilin University of Electronic Technology. His research interests include image processing, deep learning, knowledge graph, reinforcement learning and natural language.
\end{biography}

\vspace*{2.6em}
\begin{biography}[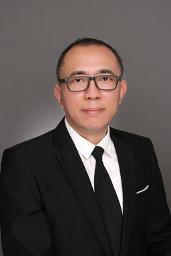]{Jing-Ye Cai} (e-mail: jycai@uestc.edu.cn) is currently a full Professor in the School of Communication and Information Engineering, University of Electronic Science and Technology of China. His research interests include communication and radar signal processing, frequency synthesis, RF and wireless systems, and spectra estimation.
\end{biography}

\end{document}